\documentclass[review]{elsarticle}
\usepackage[utf8]{inputenc}
\usepackage{graphicx}
\usepackage{subfig}
\usepackage{caption}
\usepackage{hyperref}
\newcommand*\CSSP{CLEARSY Safety Platform}

\let\today\relax
\makeatletter
\def\ps@pprintTitle{%
    \let\@oddhead\@empty
    \let\@evenhead\@empty
    \def\@oddfoot{\footnotesize\itshape
         {Submitted preprint} \hfill\today}%
    \let\@evenfoot\@oddfoot
    }
\makeatother

\begin{document}

\title{The CLEARSY Safety Platform: 5 Years of Research, Development and Deployment }
%\authorrunning{Thierry Lecomte and al.}
\author[clearsy]{Thierry Lecomte}
\ead{thierry.lecomte@clearsy.com}

\author[clearsy]{David Deharbe}
\ead{david.deharbe@clearsy.com}

\author[clearsy]{Paulin Fournier}
\ead{paulin.fournier@clearsy.com}

\author[imd]{Marcel Oliveira}
\ead{marcel@dimap.ufrn.br}

\address[clearsy]{CLEARSY, Aix en Provence, France}
\address[imd]{IMD, Universidade Federal do Rio Grande do Norte, Brazil}

%
%\titlerunning{The CLEARSY Safety Platform}  % abbreviated title (for running head)
%\author{
%Thierry Lecomte\inst{1}  \and
%David Deharbe\inst{1}  \and
%Paulin Fournier\inst{1}  \and
%Marcel Oliveira\inst{2}
%}
%
%\authorrunning{Thierry Lecomte and al.} % abbreviated author list (for running head)
%\tocauthor{Thierry Lecomte and al.}
%
%\institute{
%CLEARSY Systems Engineering, Aix en Provence, France\\
%\email{thierry.lecomte@clearsy.com} 
%\appendix
%Universidade Federal do Rio Grande do Norte, Brazil\\
%\email{marcel@dimap.ufrn.br}
%}

\begin{abstract}
The CLEARSY Safety Platform~(CSSP) was designed to ease the development of safety critical systems and to reduce the overall costs (development, deployment, and certification) under the pressure of the worldwide market. A smart combination of hardware features (double processor) and formal method (B method and code generators) was used to produce a SIL4-ready platform where safety principles are built-in and cannot be altered by the developer.
Summarizing a 5-year return of experience in the effective application in the railways, this article explains how this approach is a game-changer and tries to anticipate the future of this platform for safety critical systems. In particular, the education of future engineers and the seamless integration in existing engineering processes with the support of Domain Specific Languages are key topics for a successful deployment in other domains. DSL like Robosim to program mobile robots and relay circuits to design railway signalling systems are connected to the platform.
\end{abstract}

\begin{keyword}
formal methods \sep safety critical \sep software development \sep railway
\end{keyword}

\maketitle              % typeset the title of the contribution

\section{Introduction}
In several industrial standards (EN50128 for SIL3/SIL4, IEC61508 SIL3/SIL4, ISO 26262 for ASIL4), formal methods are highly recommended when developing safety critical software for the highest safety levels, for the specification, the development and/or the verification phases. However formal methods are highly recommended just like many other non-formal (combination of) techniques, as these recommendations are setup collectively and represent the industrial best practices. Convinced that formal methods could help to obtain better products \cite{DBLP:conf/fm/Lecomte08}\cite{DBLP:conf/fmics/Lecomte09}\cite{DBLP:conf/rssrail/Sabatier16}\cite{DBLP:conf/sbmf/LecomteDPM17}, more easily certifiable, a generic, safe execution platform has been researched for years, combining safety electronics and defect-free proven software\footnote{The software model is proved to be defect-free - complying with its formal specification and without programming errors. The code generators and the compilers are not defect-free. They are not required to be defect-free as the defects are detected with divergent behaviour during execution.}. The \CSSP\ was initially an in-house development project before being funded by the R\&D collaborative project \textit{LCHIP} (Low Cost High Integrity Platform) to obtain a generic version of the platform (i.e. not only aimed at railway systems). \textit{LCHIP}\cite{lecomte2016double} is aimed at allowing any engineer to develop a function by using its usual domain specific language~(DSL) and to obtain this function running safely on a hardware platform. With an automatic development process\footnote{The programs developed with the \CSSP\ are considerably simpler than metro automatic pilot, with few properties, simpler algorithms and hence with an expected excellent automatic proof ratio. The integration of third party provers/solvers is also expected to improve automatic proof.}, the B formal method will remain ``behind the curtain'' in order to avoid expert transactions over several languages (domain specific language, B language, interactive proof). As the safety demonstration does not require any specific feature for the input B model, it could be handwritten or the by-product of a translation process. Several DSLs are being connected (or planned to be) based on an Open API (Bxml). 

This paper demonstrates how redundant hardware and formal method can be combined to obtain a platform able to execute a safety critical application, while the developer only has to focus on the functional aspect. Hence software development may be delegated to non-expert engineers and testing is limited to validation, unit and integration testing being replaced by mathematical proof. The main contribution is cost reduction for application development, certification and deployment, as the execution platform is an order of magnitude cheaper than existing off-the-shelf solutions. It would also enable the embedding of new safety-related devices that were not previously considered because of their expensiveness.  

This paper is structured in six parts. Section 2 introduces the Terminology. Section 3 provides the rationale for designing the CLEARSY Safety Platform. Section 4 briefly introduces the B method. Section 5 introduces the architecture and safety principles of the \CSSP\, showing how the combination of the B formal method and electronic diversity ensures a high safety level, and demonstrating how the process is automated and how high profile engineers are not required anymore to complete the software development. 
Section 6 details the on-going connection with Domain Specific Languages in order to ease the \CSSP\ adoption by non-formalists.  
More abstract, section 7 shows how an Event-B model of a system could be used to derive a \CSSP\ program, allowing to increase the level of confidence on the software specification. 

%-----------------------------------------------------------------------------
\section{Terminology}
This section contains specific definitions, concepts, and abbreviations used throughout this paper.\\

\textbf{Atelier B} is an Integrated development environment (IDE) supporting the B method and the B language for software development, and Event-B for system-level analysis. Atelier CSSP is Atelier B extended with redundant code generator toolchain, bootloader, and a new project type (CSSP project).

\textbf{B0} is a subset of the B language that must be used at implementation level. It contains deterministic substitutions and concrete types. B0 definition depends on the target hardware associated to a code generator.

\textbf{Bxml} is an XML interface to B models, supported by Atelier B.

\textbf{CRC} put for cyclic redundancy check\cite{wiki:CRC}, is an error-detecting code commonly used in digital networks and storage devices to detect accidental changes to raw data.

\textbf{Diversity} refers to a method for improving the reliability of a message signal by using two or more communication channels. In our case, two diverse code generators producing different binaries from a single model allow to detect compilation errors during execution by comparing their behaviour.

\textbf{Fault tolerance} is the property that enables a system to continue operating properly in the event of the failure of some of its components. In our case, any electronic part including the processors.

\textbf{HEX} is a file format\cite{wiki:HEX} that conveys binary information in ASCII text form. It is commonly used for programming microcontrollers, EPROMs, and other types of programmable logic devices.

\textbf{PLC} put for programmable logic controller\cite{wiki:PLC}, is an industrial digital computer which has been ruggedized and adapted for the control of any activity that requires high reliability control and ease of programming and process fault diagnosis.

\textbf{Ladder logic} is a programming language\cite{wiki:LadderLogic} that represents a program by a graphical diagram based on the circuit diagrams of relay logic hardware.

\textbf{Safety} refers to the control of recognized hazards in order to achieve an acceptable level of risk.

\textbf{Safety belt} refers to the safety properties that are part of the modelling. This modelling results have to be proved against these properties. 

\textbf{Safety computer} usually refers to a computer controlling a system where the emission of an erroneous output could injure or kill people. Safety techniques (error detection, redundancy, etc.) have to be used to lower the probability of occurrence of such a failure remains below an acceptable level defined by standards.

\textbf{SIL} put for Safety Integrity Level\cite{wiki:SIL}, is a relative level of risk-reduction provided by a safety function. Its range is usually between 0 and 4, SIL4 being the most dependable and used for situations where people could die.

\textbf{Reliability} is the ability of a system to perform its required functions under stated conditions for a specified time.

\textbf{Output states (memory vs physical)} A controller computes new values for its outputs every cycle. These values (stored in memory) are used to change (with signal conversion components) the physical state of the outputs. Identity between values in memory and outputs physical state are checked regularly (with a monitor checking the signal conversion) to assess if the controller is still able to control. Depending on how the outputs are implemented (relays, transistors), changing state may take more or less time and identity check has to be delayed accordingly. 

%-----------------------------------------------------------------------------
\section{Rationale}
Developing a safety computer\cite{boulanger-safety-computer} from scratch is not something you easily decide because of the effort required to obtain such a device (several millions of Euros have been spent to develop the current CLEARSY Safety Platform). Two kinds of devices are currently available on the market for safety critical applications: PLCs (Siemens\footnote{https://new.siemens.com/global/en/products/automation/systems/industrial/plc.html}, HIMA\footnote{https://www.hima.com/en/industries-solutions/overview-of-all-hima-solutions}, etc.) and SIL3/SIL4-ready boards (MEN\footnote{https://www.duagon.com/products/computing/safe-computing-systems}, SC3\footnote{https://sc3automation.com/} etc.). PLCs provide a strict, certified environment from which it is impossible to escape, requiring systems to be designed and programmed in specific ways. On the contrary, SIL3/SIL4-ready boards offer more freedom, come with hardware features not incompatible with the standards but where the safety principles have to be fully programmed by the developer in C or similar language. 

Our first safety system development \cite{Lecomte07formalmethods} was aimed at controlling platform screen doors with a Siemens S7 PLC. The PLC was programmed with Ladder Logic\cite{franckcorbier}, one of the five programming languages mentioned by the railway standards. This PLC comes with a SIL3 certificate requiring that Ladder Logic programs must be entered by using their own internal editor. At that time, the program specification was initially written in B (and somehow loosely coupled with a preliminary system level modelling in Event-B), then implemented in B0. The B0 program was proved to comply with its specification, then translated into Ladder Logic. The resulting Ladder Logic program was manually typed in the Siemens S7 PLC. As the B0-to-ladder translator was not formally developed / verified, peer reviews took place to check the conformance of the code in the PLC with the B0 program (the code generated was very close to the B0). Because of these human activities, the development process was quite heavy and the use of formal modelling and proof was mostly nullified by these systematic reviews (many iterations were necessary to fine tune the software). Later, unfortunately (or fortunately if we consider what happened in reaction), one of our competitors managed to literally copy our system (hardware and software) and to sell it at a lower price. We then decided to develop our own solution based on the combination of redundant hardware and proven software developed with B. Producing our own hardware would reduce by an order of magnitude its cost compared to PLCs and SILx-ready boards while using Atelier B would allow more freedom and more control on the software development. The decision to go for B was easily taken as it is highly recommended by the industry standard for SIL4 software development. B is also the central formal technology we have been using during more than 20 years for most of safety critical software development. Finally the CLEARSY Safety Platform is aimed at easing the certification process, as the safety principles, embedded in the electronics design and the B software, are out of reach of the developer who cannot alter them.

Finally, the CSSP is a game changer as it proposes an alternative to the two existing safety PLCs / safety-ready boards (we are not considering in-house PLCs developed by train manufacturers which are not commercially available):
\begin{itemize}
    \item black-box PLCs with a SIL3-SIL4 stamp but also a very constrained development framework and the inability to adapt to specific requirements (signal processing on the input signal for example). The safety demonstration heavily relies on the PLC safety certificate and the bounded development environment.
    \item safety-ready boards, providing a kit of features that the developer needs to connect and activate cautiously, in time and in order, together with the application specific code. The safety demonstration has to be fully designed by the developer. 
\end{itemize}

The third way proposed by the CSSP offers the freedom to tailor the application (input/output signal processing, variable cycle time, asymmetric computing among the two processors, etc.) and to ease the safety demonstration, if the application complies with safety hypotheses coming with the certification kit. These hypotheses, named "safety application conditions", not described in this paper, have been identified during previous certification processes and are sufficient to complete the safety demonstration. Having the application B project fully proved is one of them.

%-----------------------------------------------------------------------------
\section{Introduction to the B Method}
\begin{figure}[ht]
    \centering
    \includegraphics[width=.55\textwidth]{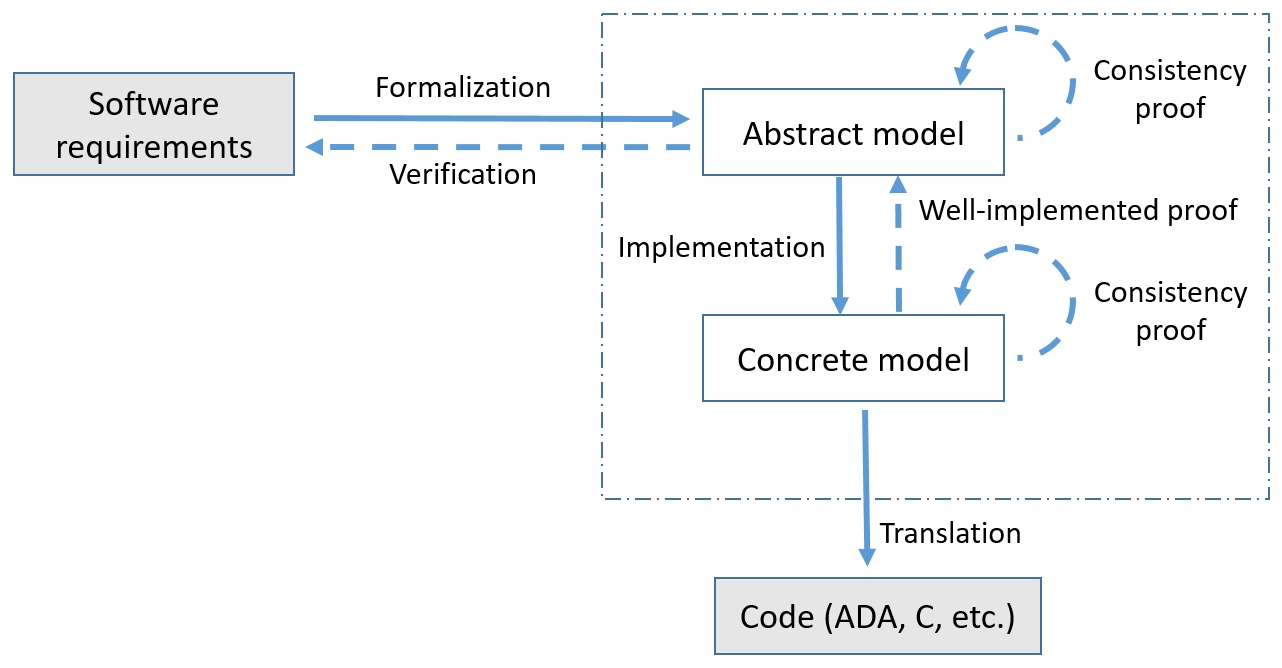}
    \caption{A typical B development cycle, from requirements to code.}
    \label{fig:cycle}
\end{figure}

B\cite{Abrial-B-Book} is a method for specifying, designing, and coding software systems. It covers central aspects of the software life cycle (Fig. \ref{fig:cycle}): the writing of the technical specification, the design by successive refinement steps and model decomposition (layered architecture), and the source code generation.  

\begin{figure}[ht]
    \centering
    \includegraphics[scale=0.4]{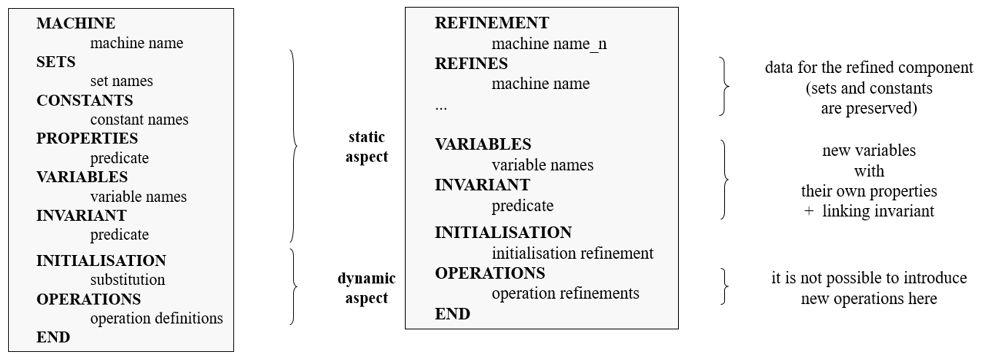}
    \caption{Structure of MACHINE and REFINEMENT components.}
    \label{fig:mach_ref}
\end{figure}

\begin{figure}[ht]
    \centering
    \includegraphics[scale=0.4]{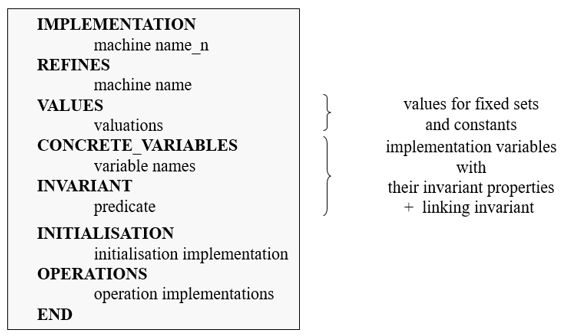}
    \caption{Structure of IMPLEMENTATION component.}
    \label{fig:implementation}
\end{figure}

B is also a modelling language that is used for both specification, refinement (Fig. \ref{fig:mach_ref}), and implementation (Fig. \ref{fig:implementation}). It relies on substitution calculus, first order logic and set theory. All modelling activities are covered by mathematical proof that finally ensures that the software system is correct. 

\begin{figure}[ht]
    \centering
    \includegraphics[width=\textwidth]{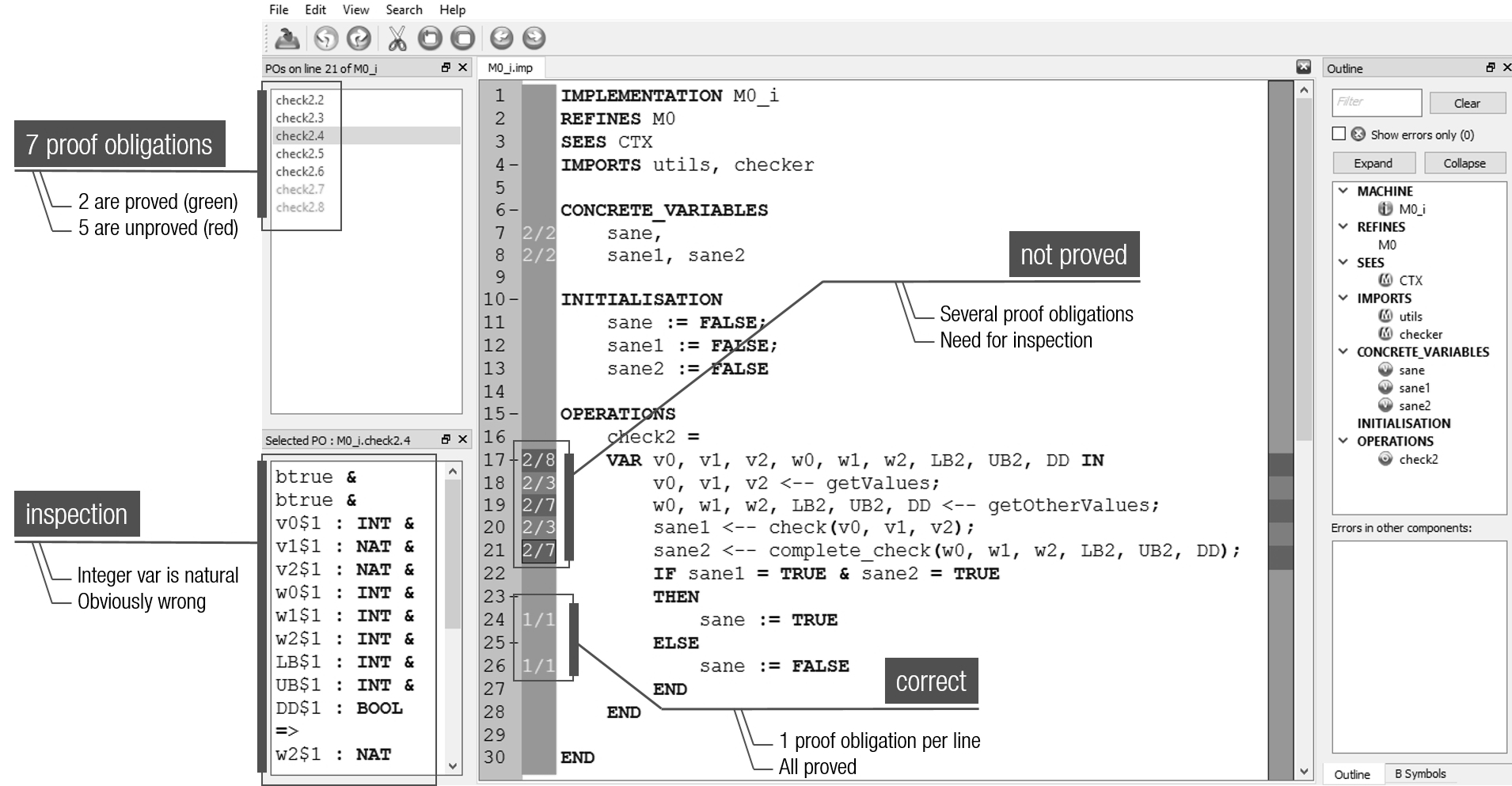}
    \caption{Atelier B model editor showing proof status.}
    \label{fig:editor-proof}
\end{figure}

B is structured with modules and refinements. A module is used to break down a large software into smaller parts. A module has a specification (called a machine) where are formalized both a static and a dynamic description of the requirements. It defines a mathematical model of the subsystem concerned:
\begin{itemize}
    \item an abstract description of its state space and possible initial states,
    \item an abstract description of operations to query or modify the state.
\end{itemize}
This model establishes the external interface for that module: every implementation will conform to this specification. Conformance is assured by proof during the formal development process.
A module specification is refined. It is re-expressed with more information: adding some requirements, refining abstract notions with more concrete notions, getting to implementable code level. Data refinement consists in introducing new variables to represent the state variables for the refined component, with their linking invariant. Algorithmic refinement consists in transforming the operations for the refined component.
A refinement may also be refined. The final refinement of a refinement column is called the implementation, it contains only B0-compliant models.
In a component (machine, refinement, or implementation), sets, constants, and variables define the state space while the invariants define the static properties for its state variables. The initialisation phase (for the state variables) and the operations (for querying or modifying the state) define the way variables are modified. From these, proof obligations are computed such as: the static properties are consistent, they are established by the initialisation, and they are preserved by all the operations. Atelier B contains a model editor merging model and proof (Fig. \ref{fig:editor-proof}) by displaying the number of proof obligations associated to any line of a B model, its current proof status (fully proved or not) and the body of the related proof obligations.

Finally a B project is a set of linked B modules. Each module is formed of components: an abstract machine (its specification), possibly some refinements and an implementation. The principal dependencies links between modules are IMPORTS links (forming a modular decomposition tree) and SEES links (read only transversal visibility). Sub-projects may be grouped into libraries. A software developed in B may integrate or may be integrated with traditionally developed code.

%-----------------------------------------------------------------------------
\section{Architecture and Safety Principles}
This section contains a description of the platform architecture and its programming model, as well as a summary of the safety principles. The safety case contains all the details leading to complete demonstration (SIL4) but are not disclosed here (the safety case is around 120 pages). The CSSP has already been certified 3 times.

\subsection{Introduction}

The CLEARSY Safety Platform is a generic PLC able to perform command and control over inputs and outputs. For safety critical applications, the PLC has to be able to determine whether it is fully functional or not. In case of failure, the PLC should move to restrictive mode where all the outputs are deactivated. The stronger the risk of harming people in case of failure, the higher the Safety Integrity Level. For SIL3 and SIL4, the computations have to be performed by a minimum of two processors and checked with a voting system. The verification listed below in Table \ref{safety:verif-dev} is used to detect PLC failures and to trigger a move to restrictive mode. 

\subsection{Architecture}

The CLEARSY Safety Platform  is made of two parts: an IDE to develop the software and an electronic board to execute this software. 

\begin{figure}[ht]
\centering\includegraphics[scale=0.2]{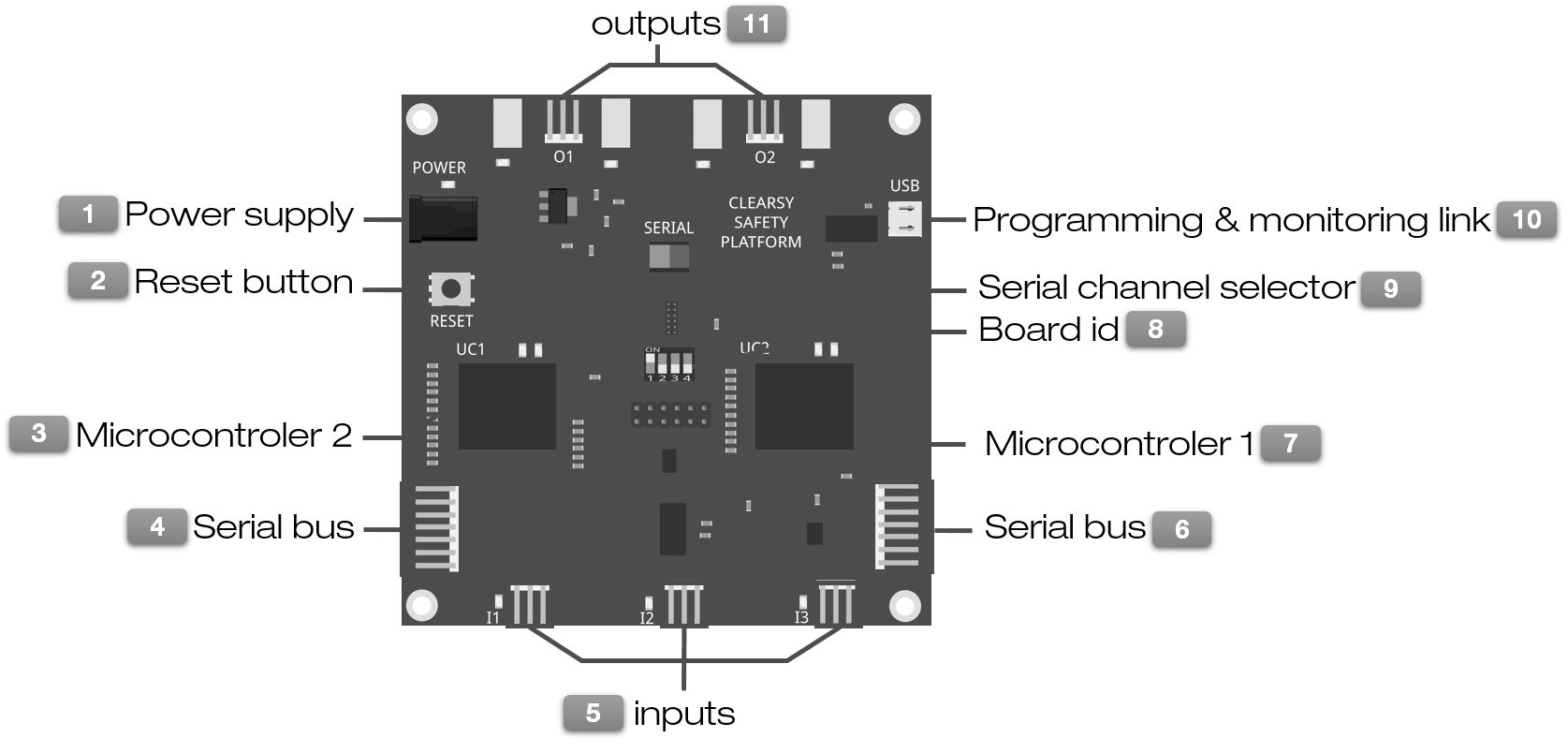}
\caption{The \CSSP\ starter kit SK$_0$ board}
\label{arch:SK0}
\end{figure}

The overall architecture (Fig. \ref{arch:SK0}) is common to all instances of the CLEARSY Safety Platform (starter kits SK$_0$ and SK$_1$). The differences lie in the number of digital (binary) IOs: 5 for SK$_0$, 28 for SK$_1$. Future instances of the CLEARSY Safety Platform will feature analog IOs and networking services (messaging) through a maintenance processor i.e. a non safety-related processor in charge of spying the microcontrollers bus and to emit traces of execution to the outside. From a safety point of view, the current architecture is valid for any kind of mono-core processor. The decision of using PIC32 microcontrollers (able to deliver around 50 DMIPS) was made based on our knowledge and experience of this processor. Implementing the CLEARSY Safety Platform on other hardware (STM32 for example) would ``only'' require to modify the existing electronic board and software tools, without much impact on the safety demonstration. Note that the PIC32 microcontrollers used by the hardware platform are commercial products. From the standards, their reliability is considered as $10^{-5}/h$.

\begin{figure}[ht]
\centering\includegraphics[width=.8\textwidth]{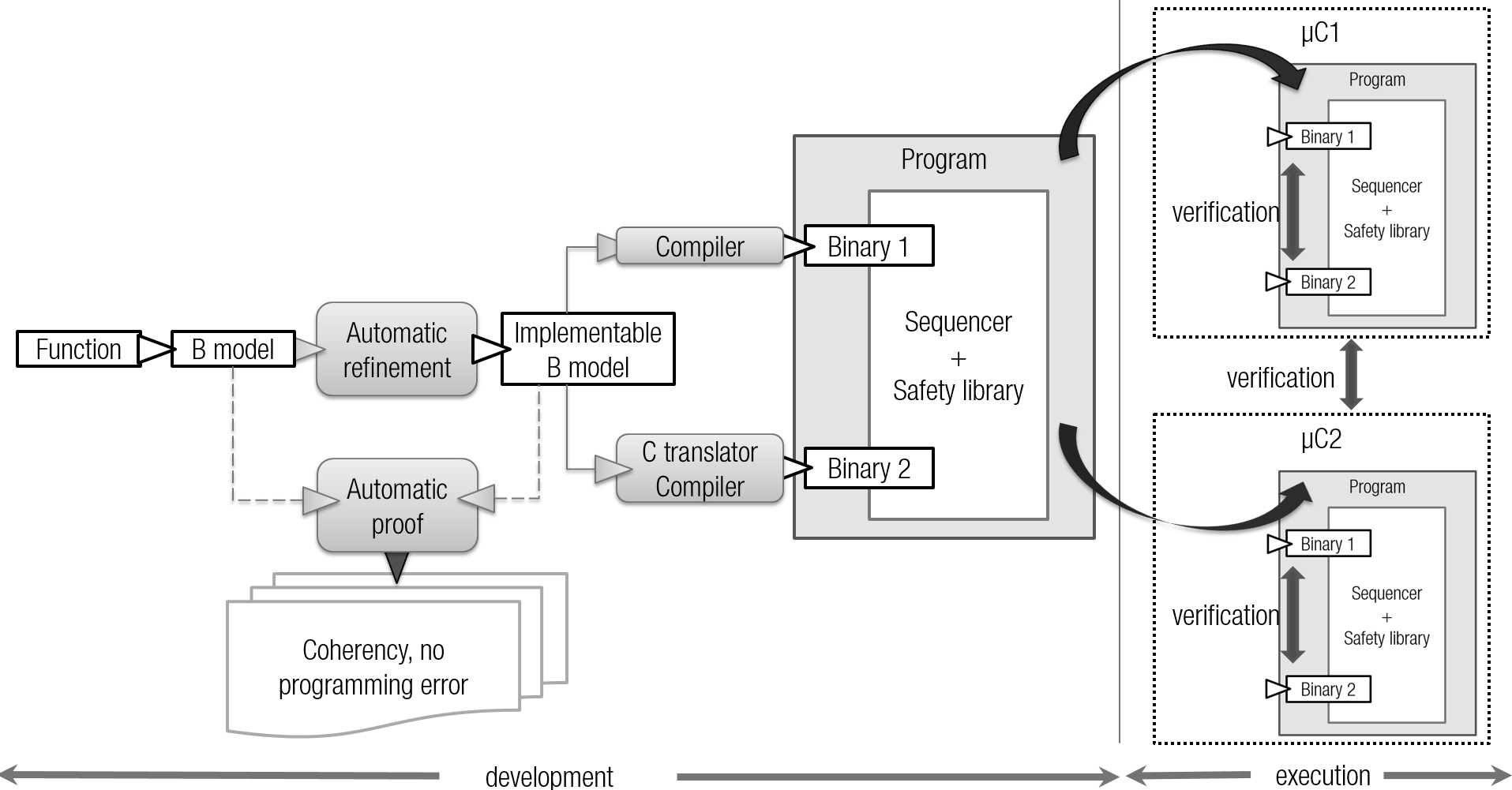}
\caption{Full path from function description to safe execution}
\label{arch:principes}
\end{figure}

The full process is described in Fig.~\ref{arch:principes} where rounded boxes are tools and rectangles are files; $\mu C_1$ and $\mu C_2$ are PIC32 microcontrollers. 

The \CSSP\ development cycle strictly follows the B method which can be summarized as:
\begin{itemize}
    \item specification model is written first from the natural language requirements (Function), then comes the implementation model, both using the same language (B). The implementable B model could be automatically refined with Atelier B BART tool, but it requires to have a fully deterministic model.
    \item models are proved to be coherent and to be correct refinements. The proof is automatic if the complexity of the model is not too high for the Atelier B theorem prover. Frequent interactive demonstrations can be turned in proof tactics to be applied automatically, 
    \item source code or binary is generated from implementation model:
    \begin{itemize}  
        \item Binary 1 (HEX file) is directly compiled from the implementation B model. The compiler has been developed in-house for supporting this technology.
        \item Binary 2 (HEX file as well) is generated in two steps. First, Implementation models are translated to C, using the Atelier B C code generator. Then the C code is compiled with gcc.
    \end{itemize}
    \item The two binaries are linked to a top-level sequencer and a safety library, both software developed in B by the CSSP IDE development team once for all, to constitute the final software.
    \item This software is then loaded on the flash memory of the two microcontrollers (bootload mode).
    \item When the board enters the execution mode or is reset, the content of the flash memory is copied in RAM for both microcontrollers which start executing it.
    \item For each microcontroller, the top-level sequencer enters a never-ending loop and
    \begin{itemize}  
        \item calls in sequence Binary 1 then Binary 2 for one iteration
        \item calls the safety library in charge of performing verification.
        \item If the verification fails, the board enters panic mode, deactivates its outputs and enters an infinite loop doing nothing.
    \end{itemize}
\end{itemize}

\subsection{Programming}

The process starts with the specification of the function to develop commonly expressed with natural language. The developer has to provide a B model of it (specification and implementation) matching the following pattern:
\begin{itemize}
    \item The function to program is a loop, where the following steps are performed repeatedly in sequence:
\begin{itemize}
    \item the inputs are read. Inputs are the same for $\mu C_1$ and $\mu C_2$, unless inputs are captured at different times, in which case the different values would cause the platform to enter panic mode.;
    \item some computation is performed in relation with the inputs/outputs status, local variables and the time elapsed since the last reset;
    \item the outputs are set.
\end{itemize}
\item The steps related to inputs and outputs are fixed and cannot be modified. 
\item Only the computation may be modified to obtain the desired behaviour.
\end{itemize}

\begin{figure}[ht]
\centering\includegraphics[width=.8\textwidth]{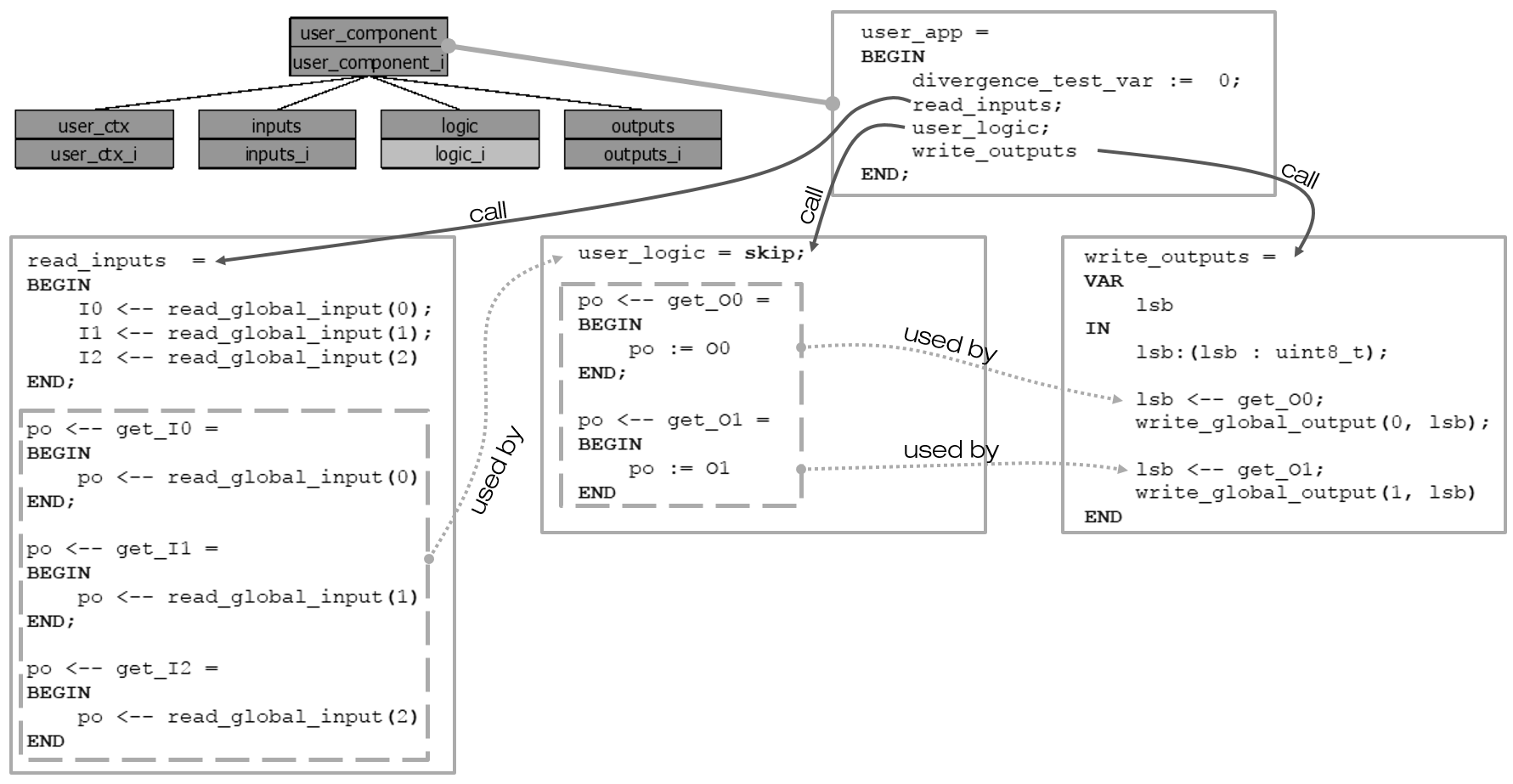}
\caption{A typical \CSSP\ project. The divergence\_test\_var is exposed as it is required by the code generation toolchain but has no impact on the modelling. }
\label{programming:all}
\end{figure}

The Atelier CSSP creates a skeleton of a B project. Fig. \ref{programming:all} shows the behavioural part of the project. The top level module is made of two components:  \textit{user\_component} (the specification) and \textit{user\_component\_i} (the implementation). This implementation imports 4 modules: \textit{user\_ctx}, \textit{inputs}, \textit{logic}, and \textit{outputs}. \textit{user\_ctx} contains only constants and sets defined by the developer; \textit{inputs} and \textit{outputs} contain variables and operations for accessing the inputs (read access) and outputs (write access). \textit{logic} contains the control \& command logic of the board. \textit{user\_ctx} and \textit{logic} are the only two modules to modify. Other modules could be required by \textit{logic} and have to be imported by \textit{logic\_i}.
Programming the board   consists of:
\begin{itemize}
    \item describing, in the component \textit{logic}, the behaviour of the board;
    \item declaring and using constants, defined in the component \textit{user\_ctx};
    \item reusing operations defined in the component \textit{inputs} and in the safety library.
\end{itemize}

\begin{figure}[ht]
\centering\includegraphics[width=.3\textwidth]{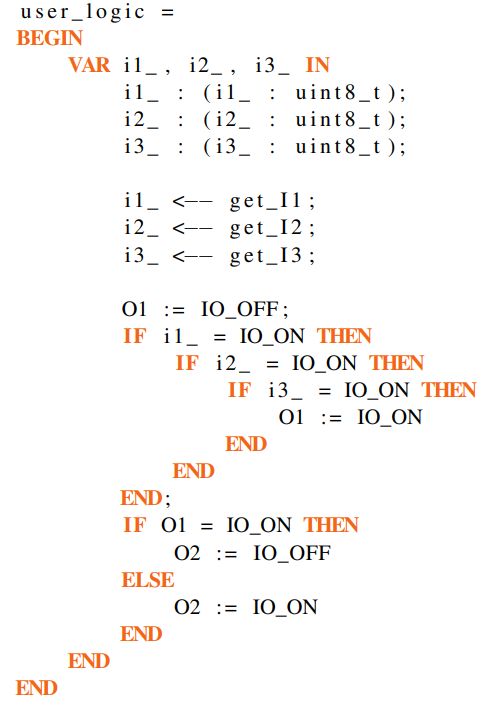}
\caption{One implementation of the operation user\_logic}
\label{programming:user_logic}
\end{figure}

The B language supported by the CLEARSY Safety Platform differs from the one described in \cite{Abrial-B-Book}, among which (Fig. \ref{programming:user_logic}):
\begin{itemize}
    \item safety variables are all unsigned integers, coded either on 8, 16, or 32 bits.
    \item digital inputs and outputs are unsigned integers coded on 8 bits (uint8\_t). Their values are either IO\_OFF or IO\_ON.
    \item testing conditions are restricted to one term only. Multiple conditions have to be in nested IF.
    \item local variables have to be typed with a "becomes such as" substitution.
    \item arithmetic operators able to produce overflow (+, -, *) are replaced by non-overflowing operators using modulo calculation to produce a result in the range of the variable receiving the result.
\end{itemize}

All the operations defined in the components \textit{inputs} and \textit{outputs} have to remain unchanged, as well as the accessing functions get\_* defined in the component \textit{logic}. Although the implementable B model, or \emph{implementation}, is usually developed manually, it may also be automatically generated with the B Automatic Refinement Tool. The B models are proved (mostly automatically as the level of abstraction of typical command \& control applications is low) to be coherent and to contain no programming error. From the implementable model, two binaries are generated:
\begin{itemize} 
    \item Binary$_1$, obtained via a dedicated compiler (developed by CLEARSY) transforming a B model into a HEX file,
    \item Binary$_2$, produced with the Atelier B C code generator, then compiled with the GCC compiler into another HEX file.
\end{itemize} 
Each binary represents the same function but is supposed to be made of different sequences of instructions because of the diversity of the tool chains. Then the two binaries Binary$_1$ and Binary$_2$ are linked with:
\begin{itemize}
    \item a sequencer, in charge of 1) reading inputs, 2) executing once Binary$_1$ then Binary$_2$, 3) setting the outputs;
    \item a safety library, in charge of performing safety verification. In case verification fails, the board enters panic mode, meaning the outputs are deactivated (no power is provided to the Normally Open (NO) outputs, so the output electric circuits are open), the board status LED starts flashing, and the board enters an infinite loop doing nothing. A hard reset (power off or reset button) is the only possibility to interrupt this panic mode.
\end{itemize}

The final program is thus made of Binary$_1$, Binary$_2$, the sequencer and the safety library. The memory mappings of Binary$_1$ and Binary$_2$ are separate. This program is then uploaded on the two micro-controllers $\mu C_1$ and $\mu C_2$. 

\subsection{Safety Principles}

For the safety case, the feared event is the wrong powering of one of the outputs i.e. this output has to be OFF (the relay should not be powered) but it is currently ON (the relay is powered). The power is provided by both microcontrollers, so if one of the two is reset, it would not power the relay and the board is in a restrictive safe state.  
The safety principles are distributed on the board and on the safety library. The safety case demonstrates that the verification performed during development and execution are sufficient to ensure the target safety integrity level.

The bootloader, on the electronic board, checks the integrity of the program (CRC, separate memory spaces). Then both microcontrollers start to execute the program. During execution, the following verifications are conducted. If any of these verification fails, the board enters the panic mode:
\begin{itemize}
    \item internal verification (performed within a single microcontroller):
\begin{itemize}
    \item every cycle, Binary$_1$ and Binary$_2$ data memory spaces (variables) are compared within each microcontroller;
    \item regularly, Binary$_1$ and Binary$_2$ program memory spaces are compared. This verification is performed ``in the background'' over thousands / millions cycles - to keep a reasonable cycle time.;
    \item regularly, the identity between memory outputs states and physical output states is checked to detect if the board is unable to command the outputs.
\end{itemize} 
\item external verification (performed between both microcontrollers):
\begin{itemize}
    \item regularly (every 50ms at the latest), data memory spaces (variables) are compared between $\mu C_1$ and $\mu C_2$. 
\end{itemize} 
\end{itemize} 

\begin{figure}[ht]
\centering\includegraphics[scale=0.3]{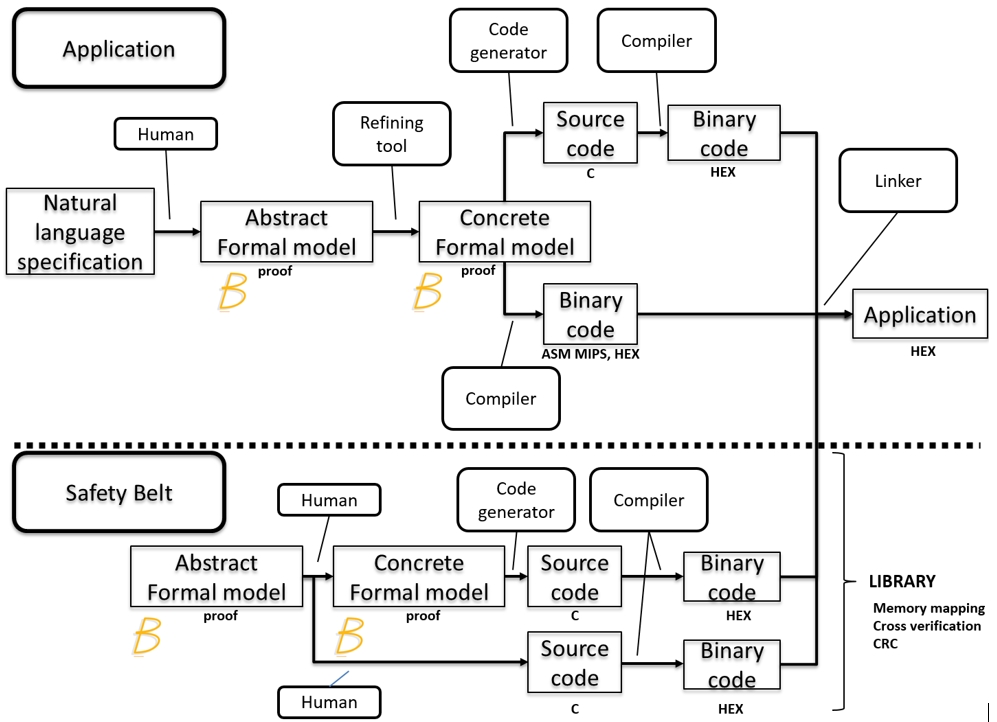}
\caption{Tools and files involved in the generation of the software}
\label{arch:process}
\end{figure}
The whole process is fully supported by dedicated tools. None of the tools part of the toolchain are proved to be correct. In Fig.~\ref{arch:process}, the tools and generated text and binary files are made explicit for both the application (the process is conducted every time an application is developed) and the safety belt (developed once for all by the IDE development team \footnote{Note that from the abstract formal model, one part of the software is developed in B with a concrete formal model, while the other part is developed manually. It happens when using B provides no added-value (for example low-level IO). A component modelled in B and implemented manually is called a basic machine.}). All the tools come from Atelier B, except:
\begin{itemize}
    \item the B to HEX compiler, initially developed to control platform screen doors for metro lines in Brazil. This tool proceeds in two steps: a translation from B to ASM MIPS, then from ASM MIPS to HEX. In order to ease debugging as ASM MIPS to HEX is a straightforward line-to-line translation.
    \item the C-to-HEX gcc compiler.
    \item the linker that combines the two HEX files with the safety sequencer and libraries.
    \item the bootloader.
\end{itemize}
Some of the tools have been ``certified by usage'' since 1998 \cite{DBLP:conf/fm/BehmBFM99}, but the newest tools of this toolchain have no history to rely on for certification. It is not a problem for railway standards as the whole product is certified (with its environment, the development and verification process, and other elements). Hence it is not required to have every tool certified. Instead the main feature used for the safety demonstration is the detection of a misbehaviour among the four instances of the function and the two microcontrollers. This way, similar bugs that could affect two independent tools at the same time and with the same effects are simply neglected: the standards incorporate the assumption that two tools developed with independent teams using different technologies could not show exactly the same buggy behaviour\footnote{Common cause failures may happen with shared conditions (same compiler, same library, same programming language, same design, same team, etc.) and break the diversity principle. If the absence of common cause failures is established during the safety analysis then the probability of occurrence of the same failure on the two paths at the same time is considered to be lower than the probability of disappearance of the atmosphere - hence neglected in the safety case.}. So a bug will always be detected by comparing the behaviour of the two tools.

\begin{table}[ht]
\caption{Verification performed during development and execution}
\small
\begin{tabular}{|l|c|l|l|}
\hline
\textbf{Stage} & \textbf{\#} &  \textbf{Failure}                  & \textbf{CSSP verification} \\ \hline
specification  & 1 & Typing error                      & Typechecker tool detects typing error \\ \hline
specification  & 2 & Specified behaviour incompatible  & Unprovable proof obligation indicates   \\
               & & with invariant properties         & specification mistake       \\ \hline           
implementation & 3 & Typing error                      & Typechecker tool detects typing error \\ \hline
implementation & 4  & Implemented behaviour incompatible  & Unprovable proof obligation indicates   \\
               &   & with invariant properties         & implementation mistake       \\ \hline           
implementation & 5 & Implemented behaviour incompatible  & Unprovable proof obligation indicate   \\
               &   & with specified behaviour          & implementation mistake       \\ \hline 
implementation & 6 & Overflow capable arithmetic operators  & Detected by the B-to-HEX compiler   \\
               &   & used instead of dedicated ones     &       \\ \hline 
implementation & 7 & IF clause with more than one condition  & Detected by the B-to-HEX compiler   \\
               &   & (B0 language restriction)          &       \\ \hline 
implementation & 8 & LOCAL variables not typed before use  & Detected by the B-to-HEX compiler   \\
               &   & (B0 language restriction)          &       \\ \hline                
code generation & 9 & Syntax errors in the C generated code  & Detected by the MICROCHIP compiler   \\  \hline
code generation & 10 & Incorrect naming in the C         & Detected by the linker   \\            
                &   & generated code                    &           \\ \hline  
code generation & 11 & Incorrect memory map             & Memory overlap detected by the   \\  
                &    &                                 &  bootloader   \\ \hline 
                compilation    & 12 & Wrong binary code generated       & Detected during execution by the safety \\
               &    &                                   & library by comparing binary$_1$ and  \\ 
               &    &                                   &  binary$_2$ variables in memory\\ 
               &    &                                   & with CRC on the same $\mu C$ \\ \hline
uploading      & 13 & Incorrect transfer between host   & Detected by bootloader during upload (CRC)\\
               &    & and electronic board              & and during execution over several cycles\\ \hline
execution      & 14 & RAM error (variables)             & Detected by comparing binary$_1$ \\
               &    &                                   & and binary$_2$ variables in memory \\
               &    &                                   & with CRC on the same $\mu C$ \\ \hline
execution      & 15 & RAM error (program)             & Detected by comparing binary$_1$ \\
               &    &                                   & and binary$_2$ program in memory \\
               &    &                                   & with CRC with the other $\mu C$ \\ \hline
execution      & 16 & Failure of one $\mu C$        & Detected by handshake between $\mu C_1$  \\ 
               &    &                               & and $\mu C_2$ at least every 50 ms \\ \hline
execution      & 17 & Outputs not command-able       & Detected by checking physical state  \\ 
               &    &                               & and command issued by the software \\
\hline
\end{tabular}
\label{safety:verif-dev}
\end{table} 

The safety is built on top of several principles:
\begin{itemize}
    \item a B formal model of the function to develop, proved to be coherent, to correctly implement its specification, and to be programming error-free i.e. no division by zero, no overflow, no access to a table outside of its range;
    \item four instances of the same function running on two micro-controllers (two per micro-controller with different binaries obtained from diverse tool-chains) and the detection of any divergent behaviour among the four instances;
    \item the deferred cross-verification of the programs on-board the two micro-controllers;
    \item outputs require both $\mu C_1$ and $\mu C_2$ to be live and running as one provides energy and the other one the command;
    \item physical output states are regularly verified to comply with the software output states, to check the ability of the board to command its outputs;
    \item input signals are continuous (0 or 5V) and are made dynamic (addition of a frequency signal) in order not to consider short-circuit current as high level (permissive) logic.
\end{itemize}
The verification performed by the CLEARSY Safety Platform, either during development or execution stages, is summarized in Table \ref{safety:verif-dev}.
   
The safety critical electronic board needs some vital elements to comply with the highest SIL requirements, such as:
 \begin{itemize}
     \item ensuring galvanic isolation between the two half-boards, to avoid that one side of the board wrongly provides energy to the other side's outputs
     \item activate safety outputs with a sinusoidal signal instead of a continuous signal, to ignore fault current. The micro-controller needs to be alive to generate the sinusoidal signal. An electrical transformer connected to the output line will generate power only if powered by alternative current.
 \end{itemize}
These features, that are not implemented on the starter kits, are only needed for real-life safety critical systems and do not prevent developers, whether students, researchers or engineers, to get educated with the CLEARSY Safety Platform and to develop prototypes.

%-----------------------------------------------------------------------------
\section{Engineering through DSLs}
The \CSSP\ is a software plant where a B model is automatically proved and transformed into a binary program that executes safely on dedicated hardware. The connection between Domain Specific Languages and the \CSSP, through a translation from DSL to B, would entitle engineers to make profit of the \CSSP\ safety features and to quite easily obtain a safety function issued from the used usual modelling language. Two experiments are being conducted: one with relay schemes for the French railways to replace wired-logic devices by programmed ones, the other with RoboSim to address the robotics domain. With these two case-studies, the objective is not to directly obtain a safety critical design but to demonstrate that the translation from these two formalisms to B (as supported by the CSSP) is doable and enables a path from one DSL model to a safe execution. In both cases, the properties expressed in the B specification are almost minimal as they are expected to have been verified at the DSL level.

\subsection{Relay Circuits \label{sec:relay-dsl}}
Relay circuits are electrical circuits that have been first considered for translation and support by the CSSP.

\subsubsection{Technical and Industrial Context \label{sec:relay-intro}}

Despite the existence of digital systems, it is still common to use relay-based systems.
Indeed, in the railway signaling domain, the interlocking systems responsible for allowing or denying trains movement are still nowadays implemented by electrical circuits containing relays. 
These electrical circuits receive, use and transmit information through the use of electromagnetic switching elements. These elements are composed of electromagnet (coils) and contacts.
The natural state of the contact (dictated by gravity) is modified by the state of the coil. 
When the coils are activated, the state of the contacts is modified thus powering or disabling other coils. Such combinations are a mean to 
implement logic functions.

\begin{figure}[htbp]
    \centering
    \resizebox{\textwidth}{!}{
    \includegraphics{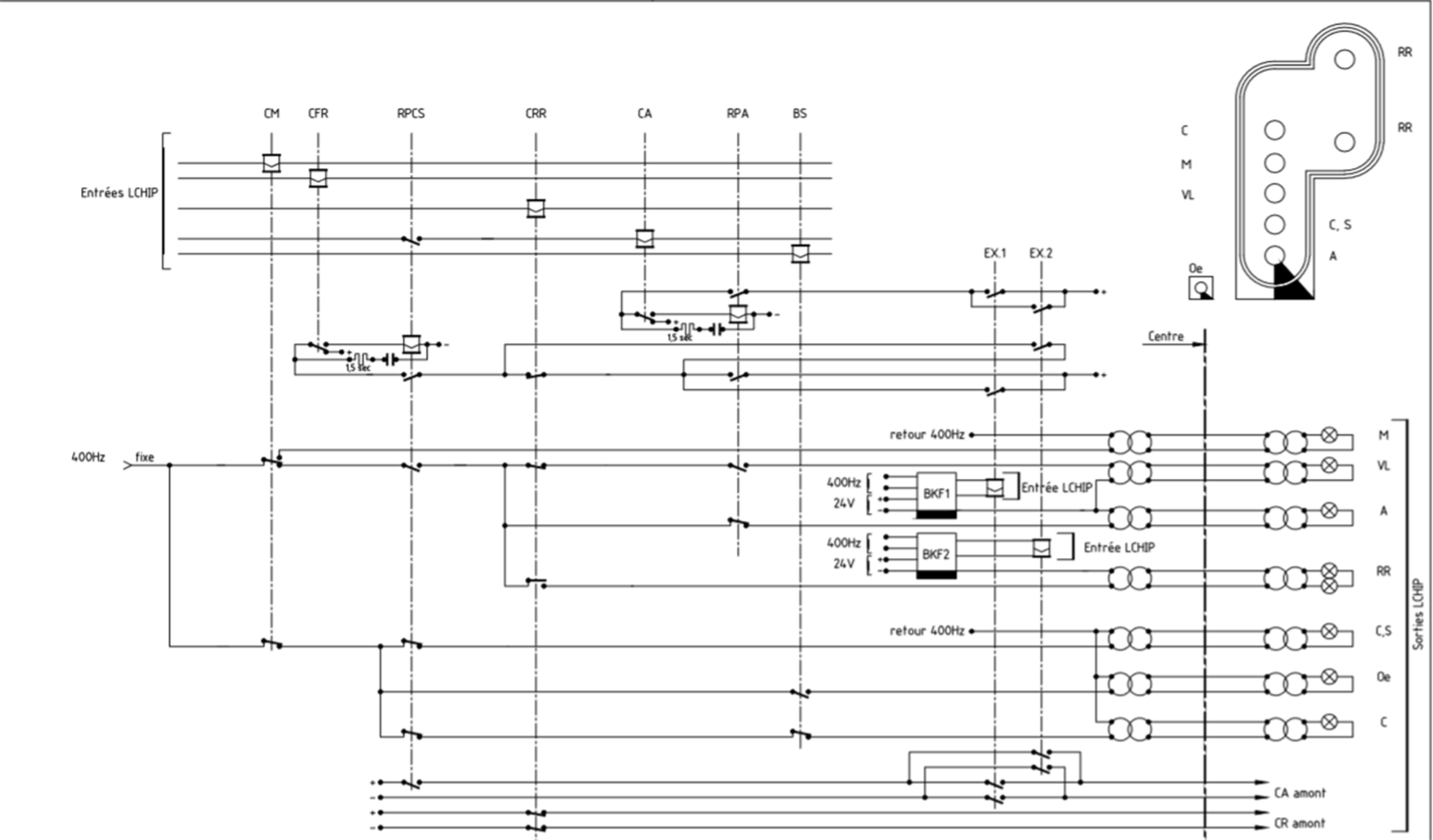}
    }
    \caption{Example of a relay circuit.}
    \label{fig:schema}
\end{figure}

Such electrical systems are designed by drawing electrical circuits, namely, relay circuits (we refer the interested reader to \emph{e.g.} \cite{retiveau}). An example of such a specification is presented in Fig.~\ref{fig:schema}. It is a circuit for controlling the state of
a color light signal installed sideways of the tracks. This design has been provided to us by \textit{SNCF Réseau\/} (owner and main manager of the French railway network).

These relay circuits are mainly composed of :
\begin{itemize}
    \item Electrical sources (positive and negative): 
    \begin{center}
        \includegraphics[scale=.5]{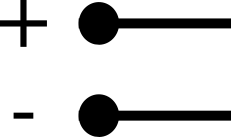}
    \end{center}
    \item Inputs, which are a special case of electrical sources that can be either activated or disabled. 
    \item Outputs, which are just electrical sources but that have effect outside of the system and thus are monitored.
    \item Mono-stable relays, which are composed of a single electromagnet; they are active when an
    electric current goes through them, inactive otherwise.
    
    In the example, the relays are depicted from left to right: CM, CFR, RPCS, CRR, CA, RPA, BS, EX1 and EX2. They are pictured with a rectangular shape containing a chevron.
    
    \item Contacts, which have two states: open or closed. A closed contact lets the current flow through whereas an open contact prevents it. Contacts are associated to a relay. 
    There are two types of contacts:
    \begin{itemize}
        \item a \emph{normally opened} contact requires its relay to be active to close;
        \item a \emph{normally closed} contact requires its relay to be active to open.
    \end{itemize}
    
    In the example, the contacts are aligned vertically with their associated relay and are
    visually connected with a dashed line. For instance, there are two contacts associated
    with the relay BS.
    The visual position of the contact indicates whether it is normally opened (drawn below the
    wire) or normally closed (drawn above the wire). For example, the top-most contact associated to
    BS is normally opened while the bottom-most is normally closed. 
    
    Notice also that, in this drawing, relay CRR is active since its normally-opened contacts
    are closed and, conversely, its normally-closed contacts are open. All the other relays
    are inactive.
\end{itemize}
Note that other components exist, such as light control block (BKF), timers, bi-stable relays and timed relays.
\subsubsection{Translating Relay Circuits Design to B Components}
To re-use the design of relay circuits to implement a safe digital circuit, 
we developed a prototype allowing to translate relay circuits to B components.
This prototype is composed of the following two steps:
\begin{enumerate}
\item Enter the circuit structure using a simple interface based on highlighting.
\item Translate automatically the entered circuit to B.
\end{enumerate}
These steps are detailed in the following. Also, since each such step may contain errors, and
since this system is critical (a wrong output may induce the train driver to enter a track section allocated to another train), it is necessary to verify their output. We also present how
these steps may be verified.

\paragraph{Highlighting}
First the user uses a graphical user interface that allows to highlight the circuit piece-wise. 
The principle is that the user must highlight every single linear wire (ending either in inputs,
outputs, electrical sources or junctions), and click sequentially on every element on the wire.
Every time the user clicks on an element, the interface queries the nature and the identity of the
component. This process allows to convert the graphical scheme to a simple tabular format.

Highlighting provides the user with a visual clue of which part of the circuit have already been
entered. Once the user has completed entering all the relevant parts of the circuit, the first
step is finished.

\begin{figure}[htbp]
    \centering
    \resizebox{\textwidth}{!}{
    \includegraphics{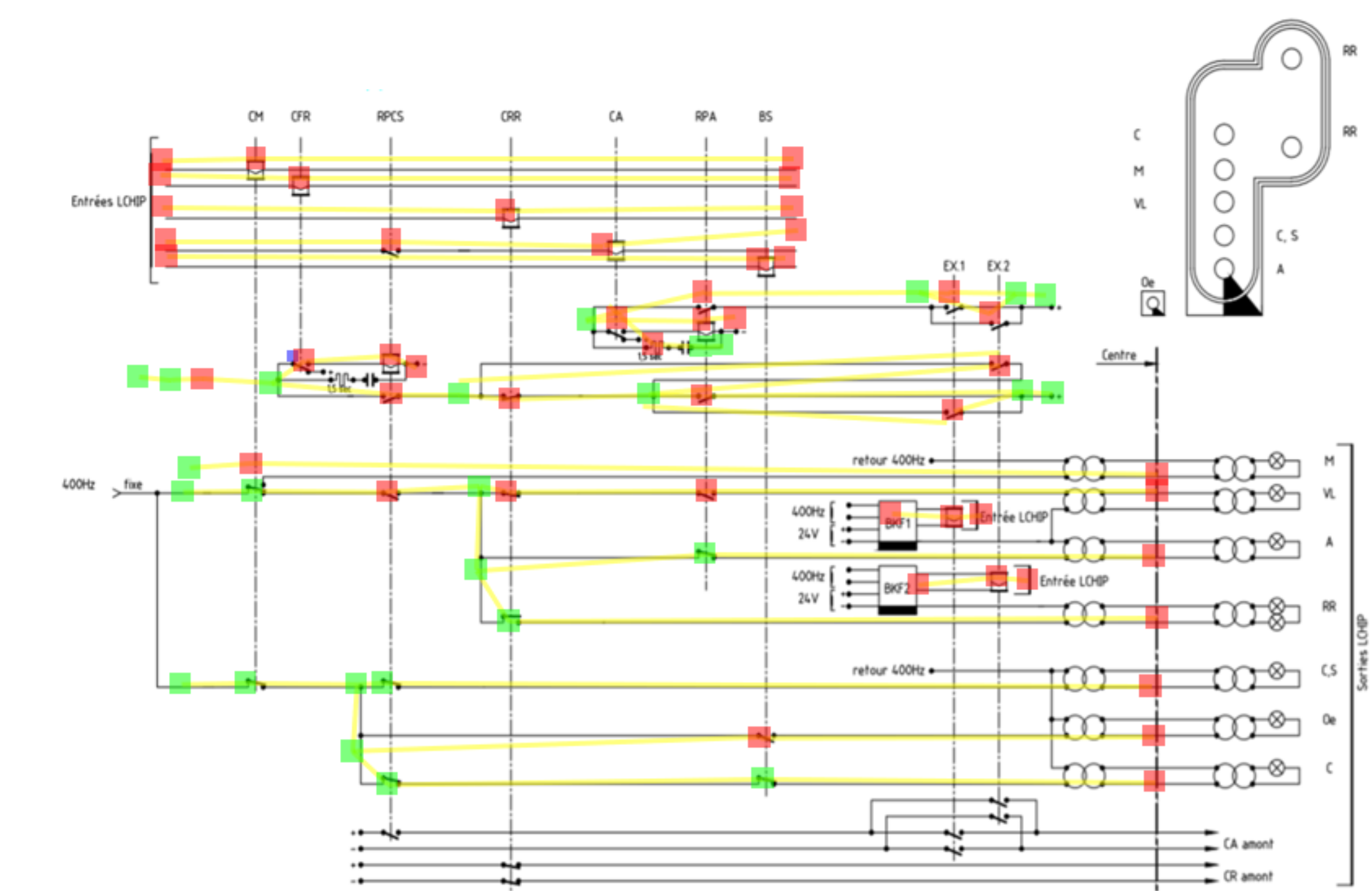}
    }
    \caption{Highlighting of the scheme of Fig. \ref{fig:schema}. Highlighted wires appear in
    yellow, entered circuit elements appear in green (when energized in default state) and red 
    (otherwise) squares}
    \label{fig:schema_highlithed}
\end{figure}

A visual representation of the result of highlighting the scheme of Fig. \ref{fig:schema} is given in Fig. \ref{fig:schema_highlithed}. Parts of the circuit that have not been entered are
greyed out. In the final setup, these parts will not be executed by the CLEARSY Safety Platform, and
they shall be implemented by other physical components.

\paragraph{Translation to B}
Once the relevant part of the circuit has been entered by highlighting, tables are filled
with all the information required to perform the translation into B components that are 
compatible with the CLEARSY Safety Platform. The principles of the translation are the following. 

The inputs of the CLEARSY safety platform are the inputs of the relay circuit, i.e., electrical sources identified as inputs by the user. Likewise, the output of the CLEARSY safety platform are
the outputs of the relay circuit.

The variables encode the states of the relays (either active or inactive). For each relay, there is
a corresponding bi-valuated variable. The computation done during a cycle of the Clearsy safety platform
consists of computing the new state of the relays according to the state of the inputs and the previous 
states of the relays. This computation is done by evaluating whether each strand is closed or open: a 
strand is open if and only if it contains at least one open contact. 
The algorithm processes each strand from a positive electrical source to a negative electrical source, 
exploring each possibility when encountering a junction. When a positive and a negative source can be 
connected with a closed strand, then the state of each relay on the strand is set to active. 
Since these new relay states may change the states of the contacts, the strands processing must be 
reevaluated. So this process is performed until a fixed-point is reached (no relay changed value
between two consecutive processing cycles).

Note that, if a fixed-point is not reached, the firmware in the CLEARSY Safety Platform guarantees 
that the system falls back to a fail-safe state. In the current version of the translator, no check is 
performed to verify whether this situation might happen. We assume that the design of the relay scheme prevents the oscillation of the electrical circuit and thus ensures the existence of a fixed-point.

Note as well that the transient states of the electrical circuit and of the Clearsy safety platform (before the fixed-point is reached) may be different. The only guarantee is that the fixed-point reached is the same as in the relay circuit. 

\subsubsection{Verification}
\paragraph{Verification of the Highlighting}
The highlighting process, being done by hand, is error-prone. We thus developed a tool to compare the result of the highlighting with the original scheme. In addition to the information needed to translate the scheme into B we also save the coordinates of the components. This allows us to re-draw a scheme from the elements highlighted and selected by the users. It is thus easy to compare (by superposition, for example) the two objects to verify that no components or connections have been omitted.

\paragraph{Verification of the translation}
Two kinds of verification may be performed for the generated B code: a \emph{structural} verification and a \emph{behavioral} verification. A structure based verification is easily achievable since the generated code has a structure that follows precisely that of the intermediate, \textit{ad hoc\/}, formats used. It would thus be easy, but cumbersome, to compare the two. We envision a tool that would implement a reverse translation from the generated B code back to the strand representation. 

Another approach is to verify the behavior of the generated B code against the expected properties of the system. For relay-based schemes, the properties express the expected values of the output after the circuit has stabilized. These properties can then be encoded in the generated B
as ASSERT instructions. This results in proof obligations generated in the B environment. Since we are dealing here with finite systems, the proof can be done by exhaustive model checking using ProB.
As an example, for a circuit commanding a light, one could require that (even in case of bulb default) the output signal is less permissive than the commanded signal. For instance, the output commanding
the green light should not be set if either the orange or the red light are commanded. 
Conversely, if the orange light is commanded, the output setting the red signal may be set, e.g., if the orange bulb is broken.

Following this approach, we have verified that the code generated for example in Fig.~\ref{fig:schema} satisfies a number of expected properties
provided together with the scheme, expressed in natural language and translated manually to first-order logic.

Model checking was conducted in a few seconds 
with ProB on this simple industrial design. It is noteworthy that, through such model checking-based verification, we have obtain a counter-example that showed a property  was not held
(this was expected by the provider of the example). It also helped us identify an early mistake in the translation process (that we have now corrected).

This approach is effective up to midsize designs (as shown by the example in Section~\ref{sec:system-analysis}) but we have not yet applied it to large circuits. However, since the translation targets a subset of the B expression language where types are essentially Booleans, alternative automatic formal verification such as SAT are also applicable and would certainly be able to address larger circuits. 

\subsubsection{Concluding Remarks}

The work presented shows the feasibility to replace heavy and physically large electrical relay circuits (they need to be fit in a cabinet sideways of the track) by smaller and cheaper digital
devices. This approach benefits from the guarantees offered by the application of formal methods
(B method, model checking) and from a generic fail-safe device (the CLEARSY Safety Platform).

\subsection{RoboSim for Robotics}

RoboSim~\cite{CSMRCDLT19} is a diagrammatic language to model simulations of robotic systems by state machines combined to define concurrent and distributed designs that use specified services of a platform. Its visual representation is akin to notations currently used by practitioners and much more friendly than any programming language. RoboSim main distinction is that its models can be verified against a UML-like design of a controller defined in RoboChart~\cite{MRLCT17}. This is possible because both notations have been given a unified semantics using CSP~\cite{Roscoe10}, a process algebra for refinement with well established tools like FDR3~\cite{fdr3}. Hence, by automatically translating their models into CSP and checking for refinement using FDR3, it is possible to automatically check correctness of simulation models regarding their design.

CSP itself has been given a UTP (Unifying Theories of Programming)~\cite{HH98} theory. This allows the encoding of the CSP semantics in the UTP making it possible to obtain support for theorem proving using the powerful prover Isabelle/HOL~\cite{FZW16}. In this context, with RoboSim, Cavalcanti et \emph{al.} fully bridge the gap between the state-machine modelling and simulation paradigms. Nevertheless, RoboSim is intended to be an intermediate notation to describe verified simulations that can be automatically translated into code for use with standard robotic simulators. In this section, we present a step towards achieving this using the CLEARSY Safety Platform.

\begin{figure}
\centering\includegraphics[scale=0.45]{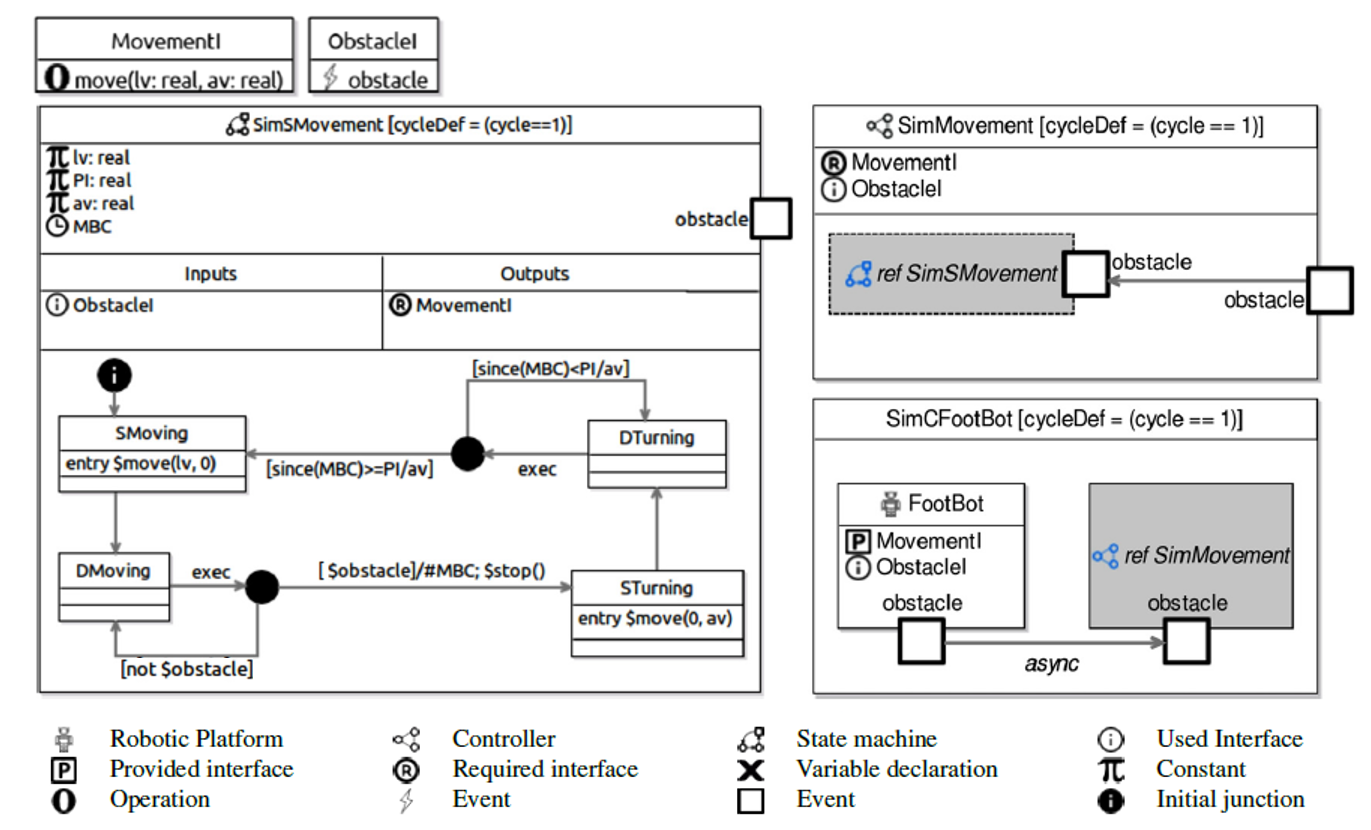}
\caption{RoboSim: obstacle detection}
\label{fig:obstacleDetection}
\end{figure}

In Fig.~\ref{fig:obstacleDetection} we present an illustrative example originally presented in~\cite{CSMRCDLT19} of a RoboSim model of a robot that can move around, detect obstacles, and stop. The module $SimCFootBot$ is composed of the robotic platform $FootBot$ and the $SimMovement$ controller that has a reference to a single simulation machine $SimSMovement$. It is important to notice that the module specifies the cycle period by including a (simple) predicate stating $cycle == 1$. The same happens with the the controller $SimMovement$ and the machine $SimSMovement$.

The interfaces can group variables, operations, and events. In Fig.~\ref{fig:obstacleDetection}, the interface $MovementI$ has the operations $move(lv,av)$ and $stop()$, provided by the robotic platform, and required by the controller. The operation $move(lv,av)$ can be used to move the robot with linear speed $lv$ and angular speed $av$. The instruction to the robot to stop is given using the operation $stop$. The interface $ObstacleI$ has just the event $obstacle$, which is used in the platform, in the controller, and in the state machine. The event $obstacle$, an abstraction of a sensor that detects obstacles, occurs when the robot gets close to any object in its environment. The robotic platform $FootBot$ defines the interface of the system with its environment via the operations of the provided interface $MovementI$ and the user interface $ObstacleI$. Assynchronously, the occurrence of the event $obstacle$ is sent to the single controller of our example $SimMovement$. The behaviour of a controller is defined by one or more state machines, specifying threads of execution. In our case, the behaviour of $SimMovement$ is defined just by the machine $SimSMovement$. It is important to notice that two different symbols denote an event:~the lighting is used when declaring an event, whereas the square is used to indicate event passing information.

State machines are similar to those in UML, except that they have a well-defined action language, and time
primitives. The state machine $SimSMovement$ has three local constants $PI$, $lv$, and $av$, and clock $MBC$. The event $obstacle$ declared in the interface $ObstacleI$ is an input, and the operations $move$ and $stop$ declared in the interface $MovementI$ are outputs.

A RoboSim model specifies a cyclic mechanism; a special marker event $exec$ defines points where
behaviour evolution must stop until the next cycle. In each cycle, inputs are read from registers, processed, outputs
are written to registers, and then time elapses in a period of quiescence until the next cycle. During processing, the simulation machine takes control of execution until progress requires the (next) occurrence of $exec$.

The visible behaviour is the reading and writing of registers, which is characterised by the inputs and outputs. Their values capture interactions corresponding to platform events, access to platform variables, and calls to platform operations. For instance, the event $obstacle$ is captured in our example as a register with a boolean value indicating whether an obstacle has been detected or not. The boolean variable $\$obstacle$ corresponding to this input is used in guards, not triggers, of transitions. In RoboSim, the only trigger used is $exec$.

The overall behaviour of $SimSMovement$ is as follows. The first cycle starts with the transition from the initial
junction to the $SMoving$ state, in which it is recorded that $move$ must be called, as indicated by $move(lv,0)$. The \$ indicates that the operation is not called immediately. Afterwards, it changes to the $DMoving$ state, where
it waits for the next cycle, because there are no transitions from $DMoving$ not triggered by $exec$. In the next cycle, $SimSMovement$ checks whether an obstacle has been perceived. If not, it remains in $DMoving$. Otherwise, it moves to $STurning$, when it resets the $MBC$ clock~(denoted by the command $\#MBC$), records that $stop$ and then $move$ must be called, besides moving to $DTurning$, all in one cycle.  In the subsequent cycle, if the amount of time since $MBC$ has been reset is less than $PI/av$, it remains in DTurning; it returns to $SMoving$ otherwise.

\subsubsection{Translation Overview}

The translation from RoboSim to the \CSSP, on which we are currently working, must consider the fact that we have two different notions of cycles. On one hand, we have the cycle of the board itself~(\CSSP\ cycle), which is able to execute around 50 million instructions per second. In each \CSSP\ cycle, the board reads the inputs from the input pins and stores their values in reserved input variables, executes the behaviour defined in an special B operation called $user\_logic$ and writes the values stored in reserved output variables to the output pins. On the other hand, we have the cycle of the simulation model (Model Cycle), which executes one cycle of its state machine possibly reading values from the reserved input variables and writing values to the reserved outputs variables. Conceptually, the time unit of the simulation does not need to be defined. Nevertheless, for execution purposes, we have to provide a definition for that. Our translation defined a constant $cycle\_unit$, which must be valuated before loading the project into the board. For the sake of our example, we assigned 100ms to the cycle unit.

\begin{figure}[ht]
\centering\includegraphics[scale=0.65]{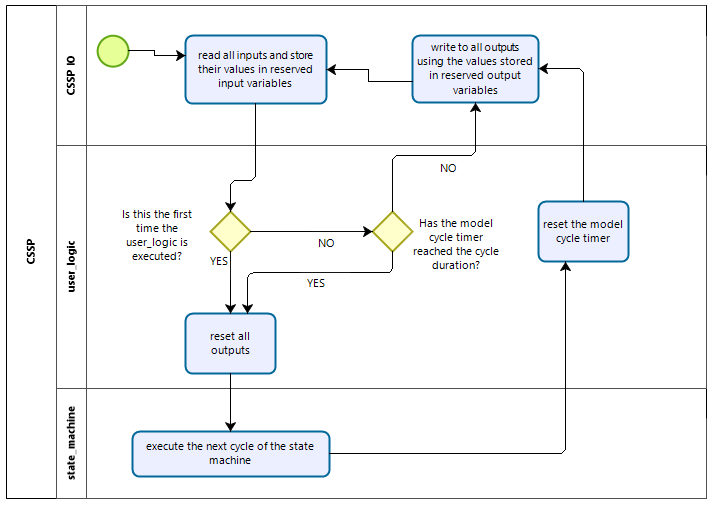}
\caption{A Summary of the resulting B implementation control flow}
\label{fig:controlFlow}
\end{figure}

A summary of the control flow of B implementation resulting from the translation of RoboSim models is presented in Fig.~\ref{fig:controlFlow}. Initially, the \CSSP\ reads all inputs from the pins and stores their values in reserved input variables. Next, we have to check if this is the first time that the $user\_logic$ is being executed. This is because RoboSim models do not wait one model cycle to provide its first outputs, which must be given immediately if the simulation model says so. For example, in Fig.~\ref{fig:obstacleDetection}, the model determines that, initially, the controller must invoke the operation $move(lv,0)$ before waiting for the next cycle~($exec$). For this reason, in the control flow presented in Fig.~\ref{fig:controlFlow}, if the $user\_logic$ is being executed for the first time, we proceed to the execution of one cycle of the controller state machine. Nevertheless, for reasons we will present later in this section, every such execution must be preceded by a reset of all outputs. Finally, we start a timer that counts the model cycle time and the \CSSP\ writes to all output pins using the values stored in reserved output variables that might have had their values changed in the execution of state machine cycle. However, if we are not executing the $user\_logic$ for the first time, we proceed to the execution of one cycle of the state machine only if the model cycle timer has reached the cycle duration. Again, we precede this execution with the outputs being reset and, afterwards, the \CSSP\ writes to all output pins using the values stored in reserved output variables. Finally, if the cycle duration has not been reached, the \CSSP\ simply writes to all output pins using the values stored in reserved output variables. In fact, the vast majority of the board cycles are empty cycles in the sense that they ignore the inputs being read and do not change any written output.

\begin{figure}%
    \centering
    \subfloat[Context Specification]{{\includegraphics[width=6cm]{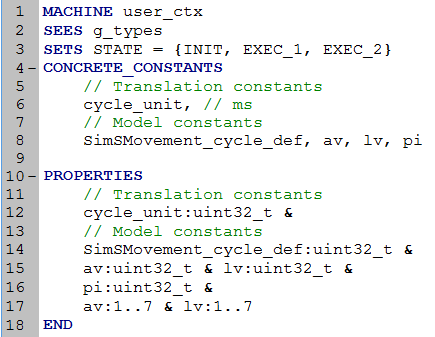} }}%
    %\qquad
    \subfloat[Context Implementation]{{\includegraphics[width=6cm]{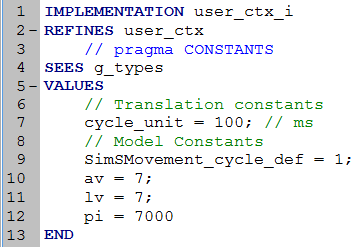} }}%
    \caption{Specification and Implementation of the Simulation Model Constants. The pragma CONSTANTS indicates a context machine impacting the safety and to be checked by the code generation toolchain. }%
    \label{fig:user_ctx}%
\end{figure}

Another important aspect is that a fine tuning of the model cycle unit is essential to make inputs noticeable by the controller and to make outputs noticeable to the robotic platform. For example, a long model cycle degrades the time between readings of the obstacle sensor and a short model cycle can make it impossible for the car engine to react to the command. Further fine tuning is also necessary in the definition of the values of each of the model constants, namely $lv$, $av$, and $pi$. All constants are specified in a separate context B machine that specifies the properties of these constants. The values of the constants are defined in an implementation B component that refines this context; hence, the B method ensures that the values assigned to all constants satisfy their properties declared in the specification. Fig.~\ref{fig:user_ctx} presents both components of our example.

\begin{figure}
\centering\includegraphics[scale=0.55]{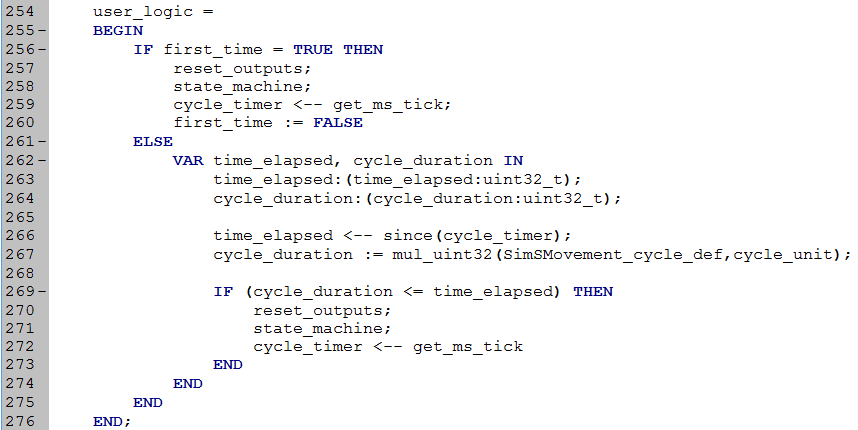}
\caption{B Implementation of the $user\_logic$}
\label{fig:user_logic}
\end{figure}

In Fig.~\ref{fig:user_logic} we present the B implementation of the $user\_logic$. In this B implementation, $first\_time$ is a state variable that is initially $TRUE$. Furthermore, $reset\_outputs$ and $state\_machine$ are operations, which set all output variables to $IO\_OFF$ and executes one cycle of the state machine, respectively.
Inputs and outputs are not coded with Boolean as a single memory perturbation is able to change one valid state to another. Hence two values have been defined, IO\_OFF and IO\_ON, both defined on 8 bits such as it is very unlikely that a memory corruption leads to the other valid state. If one output is assigned a value that is different from \{IO\_OFF, IO\_ON\} then the \CSSP\ enters panic mode.
 In order to reset the cycle timer~(lines 259 and 272), stored in the state variable $cycle\_timer$, we use the operation $get\_ms\_tick$, which gives us the current time in miliseconds. To check whether the timer has reached the model cycle duration, we compare the value of the cycle duration~($cycle\_duration$) with the time elapsed in the current cycle~($time\_elapsed$). The former is the result of multiplying~($mul\_uint32$) the constant $SimSMovement\_cycle\_def$, which is specified in Fig.~\ref{fig:user_ctx} and corresponds to the cycle duration of the $SimSMovement$ state machine defined in Fig.~\ref{fig:obstacleDetection}, with the cycle unit defined in the fine tuning of the implementation in Fig.~\ref{fig:user_ctx}, which is 100ms. The latter can be obtained using the operation $since$, which receives the value with which the cycle timer has been initialised in the last time it has been reset and returns the difference between this value and the current time, once again using the operation $get\_ms\_tick$. Finally, the operation $state\_machine$, implements the execution of the controller state machine. In our example, we have a single controller state machine. Nevertheless, RoboSim models can have many state machines with different cycle duration each. Our approach naturally deals with this possibility by using different constants for each state machine cycle duration.

\subsubsection{Translating the Controller State Machine}

In general, a translation of state machines into B is not challenging. Nevertheless, unless marked with the special marker event $exec$, RoboSim state transitions are timeless. This important characteristic would not be respected if we simply translate RoboSim models using a straightforward translation because it imposes a wait of at least one model cycle between state transitions.

\begin{figure}
\centering\includegraphics[scale=0.6]{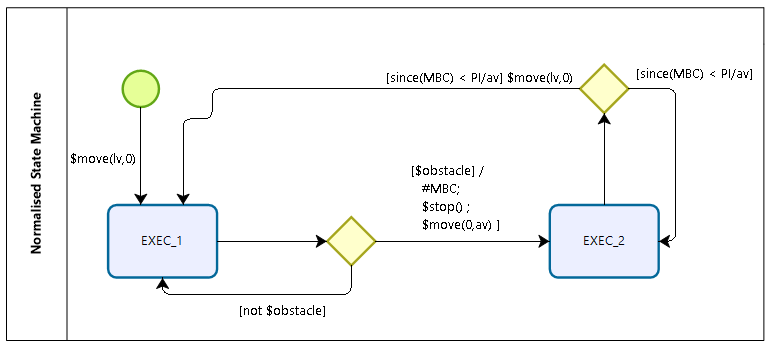}
\caption{normalized State Machine}
\label{fig:normalizedStateMachine}
\end{figure}

Our solution is to normalise the states with respect to the model cycles. The state machine resulting from this normalisation has one initial state and one state for each model cycle, which corresponds to the end of transitions marked with the special marker event $exec$. All operation calls of that cycle are composed sequentially and executed in that cycle. For instance, in Fig.~\ref{fig:normalizedStateMachine} we present the result of normalising the state machine of Fig.~\ref{fig:obstacleDetection}.

In the normalized state machine we only have three states: \begin{itemize}
    \item $INIT$: corresponds to the end of the transition leaving the initial junction. In Fig.~\ref{fig:normalizedStateMachine} represented with the circle;
    \item $EXEC\_1$: corresponds to the end of the transition leaving the state $DMoving$;
    \item $EXEC\_2$: corresponds to the end of the transition leaving the state $DTurning$.
\end{itemize}
These states are specified as members of an enumerated set, $STATE$, which is declared in the context machine presented in Fig.~\ref{fig:user_ctx}.

Now, the translation of the normalized state machine is relatively trivial. Initially, the code invokes operation $move(lv,0)$ and enters state $EXEC_1$. In the remaining execution, the resulting code always waits one model cycle before leaving the current state. The main differences of the normalized state machine of our example with that presented in Fig.~\ref{fig:obstacleDetection} are:

\begin{enumerate}
    \item operation calls placed in states, like $\$move(lv,0)$ originally in state $SMoving$, are now in the state transitions, and
    \item the transition from $EXEC_1$ to $EXEC_2$, in which the commands $\#MBC; \$stop()$, originally in the transition from $DMoving$ to $STurning$, and $\$move(0,av)$, originally in the entry of state $STurning$, are sequentially composed in a single transition.
\end{enumerate}

\begin{figure}
\centering\includegraphics[scale=0.55]{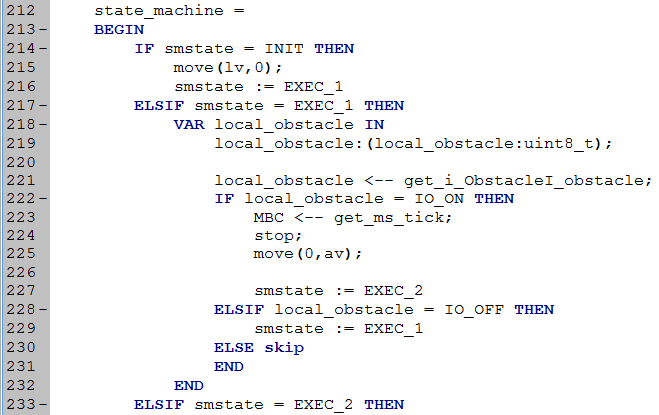}
\caption{B Implementation of the State Machine}
\label{fig:state_machine}
\end{figure}

In Fig.~\ref{fig:state_machine} we present part of the B implementation of the state machine of our example. All source files can be found at \url{http://bit.ly/2JtkxuQ}, where you can find the complete Atelier-B project of our running example and an Arduino program that emulates the behaviour of the robotic platform\footnote{After creating the \CSSP\ project, the only files we have edited are $RoboSim\_ctx$, $RoboSim\_ctx\_i$, $logic$ and $logic\_i$~(the main file, in which all operations mentioned in this paper can be found)}. The $smstate$ is a state variable that is initialised with $INIT$. Hence, the first time this operation is invoked, this machine invokes the operation $move(lv,0)$ and updates the $smstate$ variable to $EXEC_1$. The control returns to the operation $user\_logic$, which only invokes the $state\_machine$ after it reaches the cycle duration. Now, this operation uses the \CSSP\ operation $get\_i\_ObstacleI\_obstacle$ to get the value of the reserved input variable, $obstacle$. As a standard, we prefix the name of all inputs like $obstacle$ with an $i$ and the name of its interface. For example, $i\_ObstacleI\_obstacle$ corresponds to the the input signal $obstacle$ of the interface $ObstacleI$. If the value retrieved is $IO\_ON$, the  clock $MBC$, implemented as a state variable, is reset~(line 223), the operations $stop$ and $move(0,av)$ are invoked in this order~(lines 224 and 225) and the state machine remains in the current state, $EXEC\_1$.

\subsubsection{Translating Operation Calls}

As previously presented, the operation calls of RoboSim models are directly translated to the invocation of operations of the B implementation. In order to follow the RoboSim semantics presented in~\cite{CSMRCDLT19}, we need to consider, for each operation of a RoboSim model, a boolean output value and the operation output values. The former indicates that the operation has been invoked in the current cycle.    

\begin{figure}
\centering\includegraphics[scale=0.55]{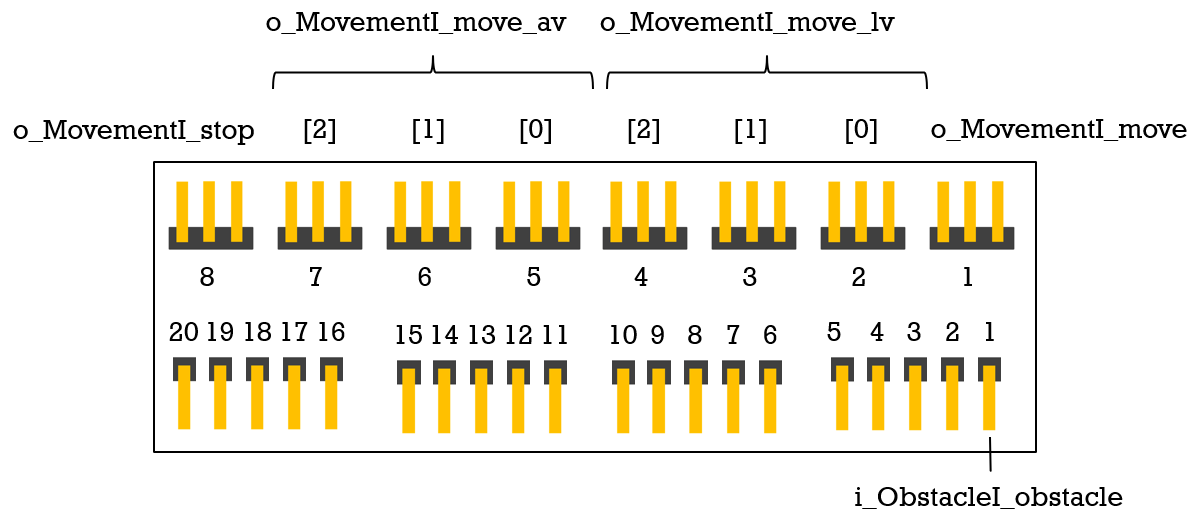}
\caption{\CSSP\ Inputs and Outputs}
\label{fig:cssp_pins}
\end{figure}

An important restriction is the number of input and output pins available in the \CSSP. In its current version, SK1, the board provides 20 inputs and 8 outputs. During the project creation, we configure the \CSSP\ board by mapping each pin to the corresponding input/output. Fig.~\ref{fig:cssp_pins} presents the mapping we have implemented in our example. The first input pin is used to receive the only input, $obstacle$. Our translation uses one output pin for each output operation to indicate that it has been invoked:~the output 1 indicates that $move$ has been invoked and output 8 indicates that $stop$ has been invoked. Similarly to the inputs, our standard prefixes the name of all outputs like $move$ with an $o$ and the name of its interface. For example, $o\_MovementI\_move$ corresponds to the invocation of the output operation $move$ from the interface $MovementI$. Finally, we are left with six output pins which are used to output the values of $lv$~(pins 2, 3 and 4) and $av$~(pins 5, 6 and 7). For output arguments, we use the name of the argument and the index of the bit as suffixes. For example, $o\_MovementI\_move\_lv\_0$ corresponds to the least significant bit of the argument $lv$ of the operation $move$. 

The limitation on the number and type of outputs imposes a property of the constants used in the model. Both, $lv$ and $av$, can only receive natural values ranging from $0$ to $7$. This platform restriction is included in the $PROPERTIES$ clause of the context machine presented in Fig.~\ref{fig:user_ctx}, in which we include the predicate $av:0..7~\&~lv:0..7$. As for all other constants, the B method ensures that the values assigned to these constants in the B implementation satisfy these properties.   

\begin{figure}
\centering\includegraphics[scale=0.55]{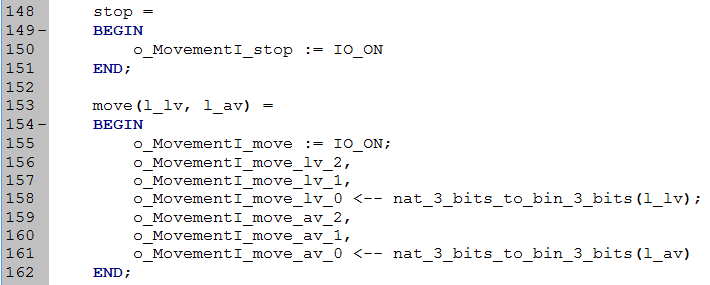}
\caption{Operations $move$ and $stop$}
\label{fig:local_operations}
\end{figure}

By way of illustration, in Fig.~\ref{fig:local_operations}, we present the implementation of the operations $stop$ and $move$. The former implements a parameterless model operation; hence, it simply indicates that the operation has been invoked by assingning $IO\_ON$ to the reserved output variable that corresponds to the operation $stop$, $o\_MovementI\_stop$~(line 150). The latter, however, implements a model operation with arguments. For this reason, besides indicating that the operation has been invoked~(line 155) it also assigns the values of the arguments $lv$ and $av$ to the corresponding output pins~(lines 156 to 161). A local operation $nat\_3\_bits\_to\_bin\_3\_bits$ is used to convert the natural number ranging from 0 to 7 into a binary number and assigns each of its bits to the right corresponding pin. 

The translation from RoboSim to the \CSSP\ has proved to be an interesting subject and application of \CSSP\ for robotic platforms. We are currently working on more elaborated and complex simulation models that will validate our current translation strategy and raise the need for more complex solutions. For instance, some of the models that are in our translation plans have more than 8 outputs (SK$_1$ board). Nevertheless, it is possible to connect different boards in sequence in a way that some of the outputs of one board are inputs to a different board. An investigation on how the model behaviour can spread among different boards is in our near future research agenda. A crucial result that will allow the application of our approach in industry is the automation of our translation strategy. This implementation is also in our research agenda and will define the level of user interaction in the translation process. For example, most of the simulation models like our example model, as expected of simulation models, do not define constant values and cycle unit. This, however, is essential to execute the resulting program in the \CSSP\ and needs to be given at some point by the user to the translator. 

Finally, as for RoboChart and RoboSim, we intend to provide a CSP semantics to our B implementations. By doing so, it will be possible to automatically check correctness of our B implementations regarding their simulation models by checking for refinement using FDR3. Furthermore, this also allows the encoding of the CSP semantics in the UTP making it possible to obtain support for theorem proving using the powerful prover Isabelle/HOL~\cite{FZW16}.

%-----------------------------------------------------------------------------
\section{Applications from a System-Level Formal Analysis
\label{sec:system-analysis}}
This section presents how the design of a CSSP-based product may be conducted in a process that 
originates in the formal analysis of a system design and proceeds with a model-based design 
realized through decomposition and refinement. 

\begin{figure}[ht]
    \centering
    \includegraphics[width=.5\textwidth]{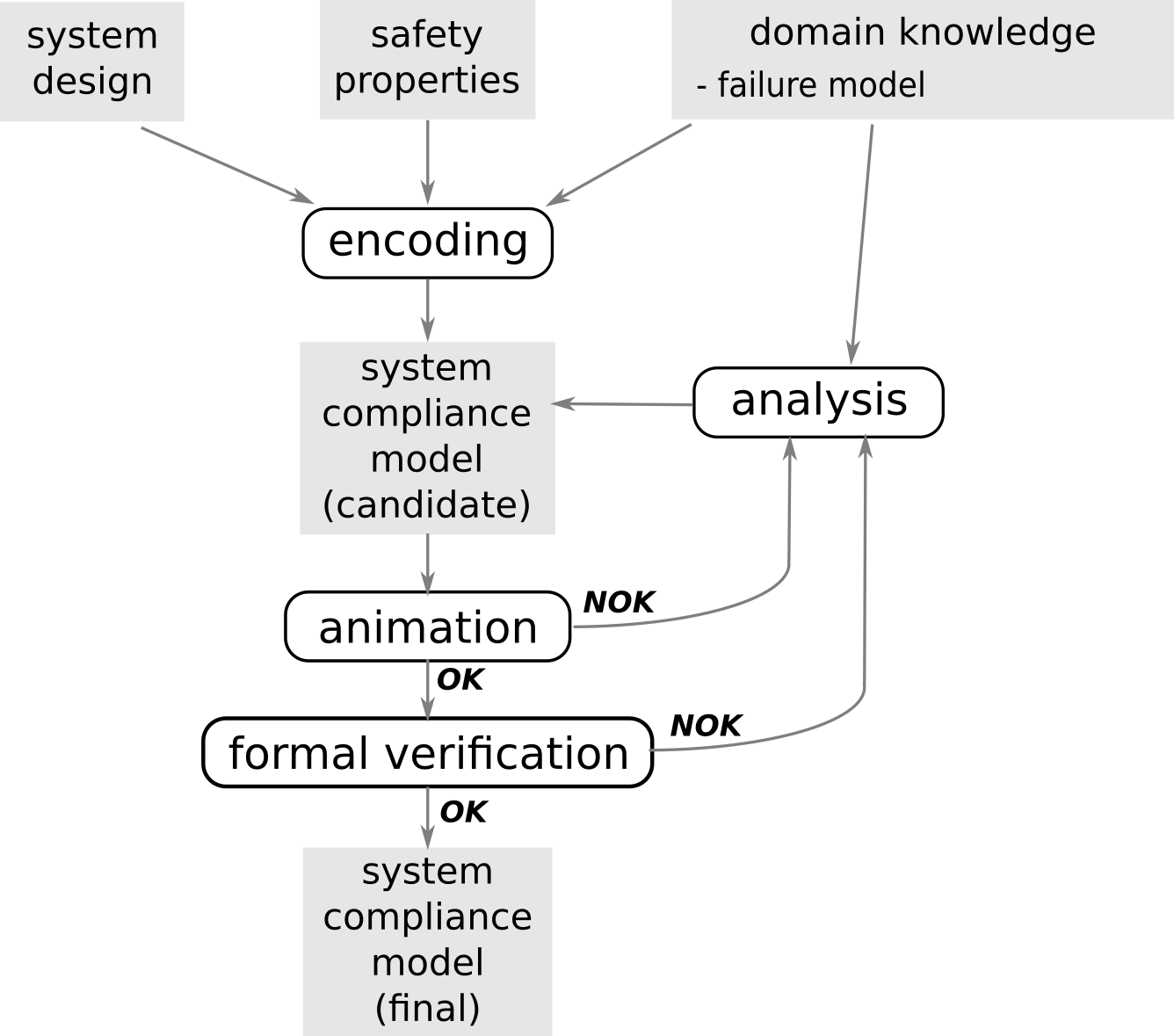}
    \caption{The formal analysis process}
    \label{fig:formal-analysis-process}
\end{figure}

The inputs of this process are:
\begin{itemize}
    \item a system design;
    \item one or several safety properties that must be ensured by the system;
    \item domain knowledge;
    \item identification of the elements of the system that should be implemented in a CSSP-based board;
\end{itemize} 

The formal analysis is based on the first three inputs. Once the formal analysis has been conducted, the
last input is used to derive a specification for the CSSP-based component that will be part of the system. 

The output of the analysis is a formal model that contains not only the system design logic, but also the 
required safety properties and all the hypotheses that are necessary to ensure that the system
meets the properties. These hypotheses synthesize the domain elements that are
necessary to establish the demonstration that the design complies to the safety requirements. These
hypotheses must be validated by domain experts. In the case of a fail-safe design, part of the domain 
knowledge is the possible failure modes of the devices used to implement the system design.

The formal analysis process is pictured in Fig.~\ref{fig:formal-analysis-process}. The process 
initiates with the construction of a model that encodes the system design, the
required safety properties (and possibly the failures). The model is first animated to ensure that
it matches the expected behavior. The feedback from this animation might be to include some 
domain knowledge into the model. A typical domain-oriented constraint would be that a fail-safe
sensor does not miss any event that it is supposed to detect. Formal verification is also
applied, either through systematic exploration (model checking) or reasoning (proof).

\begin{figure}[ht]
    \centering
    \includegraphics[width=.8\textwidth]{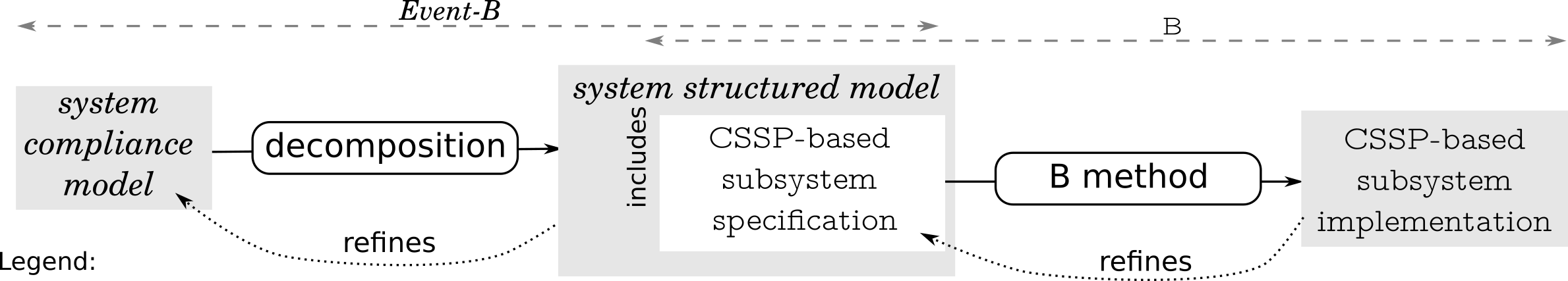}
    \caption{From a System-Level Compliance Model to a CSSP-Based Solution.}
    \label{fig:component-design}
\end{figure}

Once the system compliance model has been constructed, the next step is to produce the specification of
the system component to be implemented with a \CSSP\ board (see Fig.~\ref{fig:component-design}). This
step is performed using both Event-B~\cite{Abrial-Event-B} and the B method.
This step takes as input the system compliance
model produced by formal analysis, and the boundaries of the sub-system that shall be implemented
as a CSSP-based product. These inputs guide the decomposition of the compliance model into a structured
model of the system. In this model, all the logic that is to be executed by the CSSP-based subsystem 
is factored into a B machine. This decomposition is conducted using the Event-B modeling language,
extended to allow component structuring constructions. 

In the realm of the B method, the machine thus obtained is the \emph{specification} of the function to 
be executed by the \CSSP. It is the starting point of the refinement based design approach of the
B method, which we use to obtain an implementation capable of being compiled and uploaded to a \CSSP\ board (see fig~\ref{arch:process}).

\begin{figure}[ht]
    \centering
    \includegraphics[width=.8\textwidth]{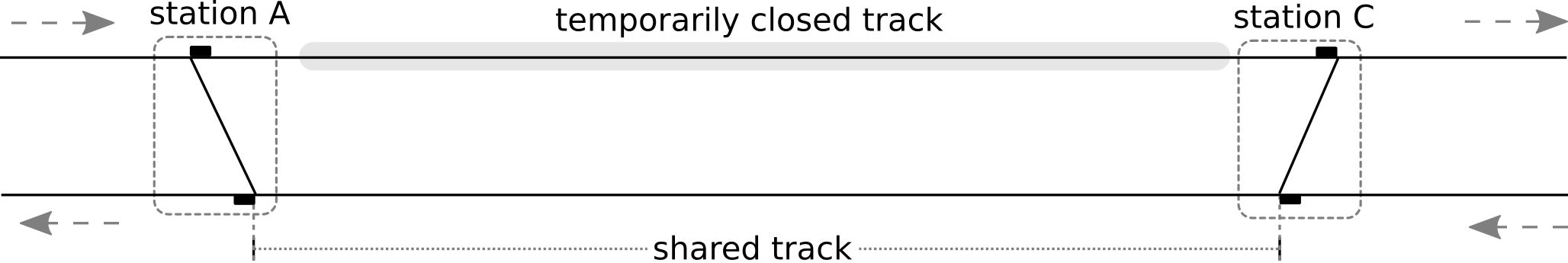}
    \caption{Provisional Wrong-Way Installation}
    \label{fig:itcs}
\end{figure}

We have applied this approach for an interlocking system for the railways: a temporary wrong-way
interlocking (see Fig.~\ref{fig:itcs}). This is a system that is used to manage a track that is temporarily
shared between
two lines, when a portion of one line is to be temporarily closed. It is composed of two 
temporary stations, A and C, located at each end of the shared track portion, as well as several
sideways equipment, fixed and temporary. 
The safety property is the absence of front collision on the track. 
The given system design was a relay-based solution for both stations A and C (presented in~\cite{DBLP:conf/rssrail/PereiraDPB19}).
We applied the approach described in this section to build a B module implementing most elements composing station A and derived
an implementation compatible for a CSSP-based solution. In practice this would allow to safely replace the
expensive, heavy cabinet of interconnected relays, that has to be installed sideways the track nowadays,
with a much smaller and lighter fail-save electronic device based on the \CSSP.

\section{Conclusion and Perspectives}
\paragraph{Exploitation}
The \CSSP, combined with improved proof performance and connection with Domain Specific Languages, pave the way to easier development of SIL4 functions (including both hardware and software). The platform safety being out of reach of the software developer, the automation of the redundant binary code generation process and the certificates already obtained for products embedding \CSSP\ building blocks, would enable the repetition of similar performances without requiring highly qualified engineers. The CLEARSY Safety Platform building blocks have been used in successive projects where these building blocks have been modified / improved to fulfill diverse requirements. Even if complete cost reduction figures are not yet available, our findings are that software development and certification are reduced by at least 30\% as the safety principles do not need to be designed/programmed and as a significant part of the safety case comes from the certification kit (a set of documents explaining how the CLEARSY Safety Platform safety was designed, implemented, tested, and verified, and how the CLEARSY Safety Platform has to be integrated into target hardware - the so-called exported constraints).
Moreover, the hardware platform is generic enough to host a large number of complexity-bounded industry applications, with a special focus on the robotics and autonomous vehicles/systems domains. Intelligent road infrastructure also seems of interest, as it appears that fully autonomous cars would require additional support from their environment to deliver a really safe mobility service. This aspect is going to be developed in the coming years.

\paragraph{Dissemination}
The CSSP IDE is based on Atelier B 4.5.3, providing a simplified process-oriented GUI. It also contains the toolchain to generate the binary, and a bootloader to upload the binary produced on the CSSP board. A first starter kit, SK$_0$, containing the IDE and the execution platform, was released by the end of 2017\footnote{https://www.clearsy.com/en/our-tools/clearsy-safety-platform/}, presented and experimented at the occasion of several hands-on sessions organized at university sites in Europe, North and South America. Audience was diverse, ranging from automation to embedded systems, mechatronics, computer science and formal methods. Results obtained are very encouraging: 
\begin{itemize}
    \item Teaching formal methods is eased as students are able to see their model running in and interacting with the physical world. It was the occasion to demonstrate how formal methods could be used with embedded systems and IoT. Fruitful discussions took place about how to specify / guarantee performances, what can or cannot be proved with such systems, etc.
    
    \item Less theoretic profiles (computer science, mechatronics, automation) may be introduced/educated to more abstract aspects of computation. \textit{clock} and \textit{combinatorial} exercises were a starting point for specification enrichment and the discovery of the formal proof. Of course, the pedagogical objective in term of formalization was lower than with more formal profiles, but the students managed to understand the absence of programming error and the non-deterministic substitutions for simple modelling.   
    
    \item The platform has demonstrated a certain robustness during all these manipulations and has been enriched with the feedback collected so far. Several electronics / software errors were detected during the preparation of course when designing exercises, others during these exercises:
    \begin{itemize}
        \item USB interface is used to program the board and to power it. The second release of the board embeds LEDs to show inputs and outputs status. Many computers do not provide enough current to power all the LEDs, leading to erratic behaviour. The workaround is to power the board with a power supply instead of the USB cable.
        \item Time synchronization algorithm between microcontrollers was erroneous. It was not detected during short programming sessions but after leaving boards running during (quite long) coffee breaks.
    \end{itemize}

    \item The IDE GUI was improved with the automation of the code generation process and the display of a carousel showing graphically the progress of the generation. The configuration of the board was also simplified, by displaying the position of the switches on the board and by filling the configuration file with default inputs and outputs names.
    
    \item \CSSP\ is yet used to teach in Master 2 in universities and engineering schools. Electronic documentation\footnote{Available at https://www.clearsy.com/en/our-tools/clearsy-safety-platform/download-clearsy-safety-platform/} is used to structure the courses and is updated every 2 months. With three inputs and two outputs, the starter kit SK0 is for discovering the technology; another version of the board is planned for 2020 able to handle more I/O (up to 64).
\end{itemize}

\paragraph{Future}
The \CSSP\ is a software plant able to generate automatically software for safety critical applications and guarantee its safe execution (outputs are deactivated in case of misbehaviour). This way, it is not required that the developer knows (and masters) all the technical details of the design.

\begin{figure}[ht]
    \centering
    \includegraphics[width=1.\textwidth]{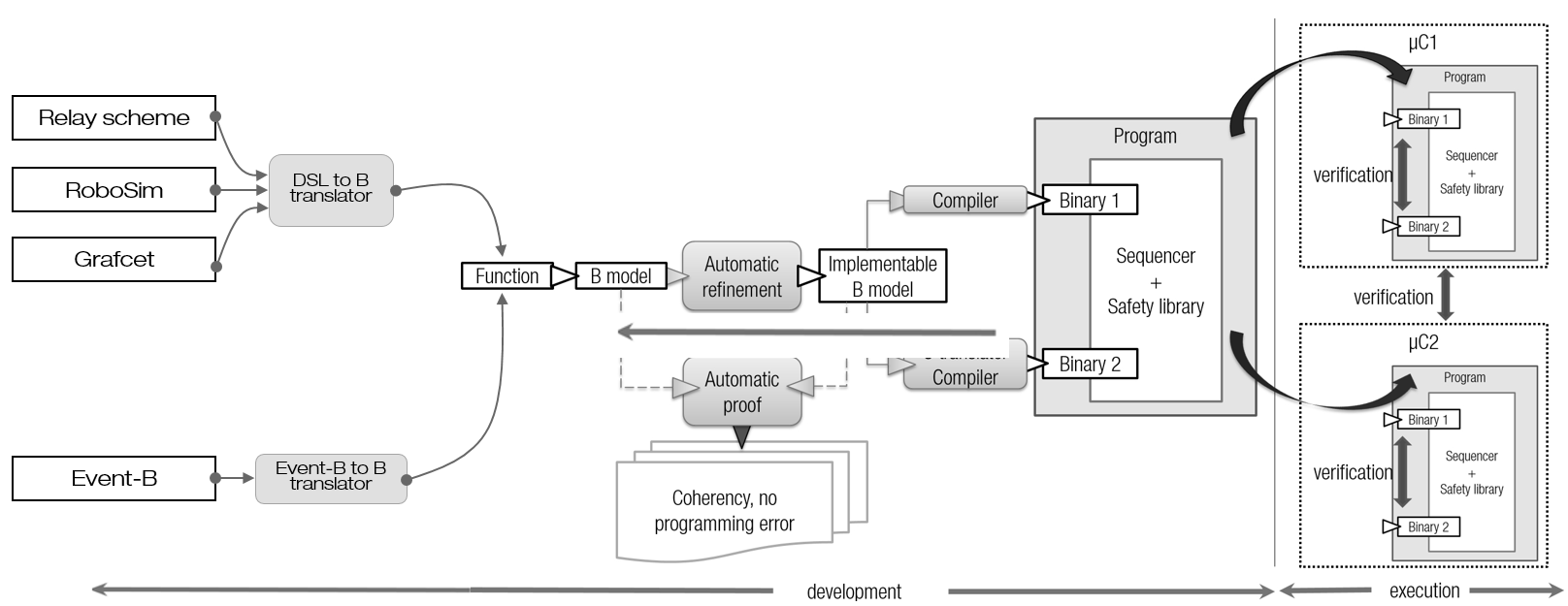}
    \caption{The complete picture including connection with DSLs and system-level proven models. The connection with Grafcet, ongoing, is required to connect with PLCs}
    \label{process-future}
\end{figure}

Moreover, the connection of the \CSSP\ with domain specific languages, expected to fully hide the formalities, does not perturb the developer in his design activities. The possibility to derive \CSSP\ software specification from a proven system-level specification improves the level of confidence of the final system. Finally the \CSSP\ building blocks have been embedded and certified in a number of railway projects in Brazil, Sweden and US, with diverse certification bodies. The \CSSP\ is expected to lower the cost of certified safety systems in a number of industrial domains, to contribute to increase citizens safety in our always-more-automated world, and also to convert students and engineers to formal methods due to its ease of implementation.

\paragraph{Limits of the approach}
The \CSSP\ is an innovation combining a number of existing results, many of them issued from previous completed software and electronic projects at CLEARSY. The core of the \CSSP\ (software toolchain, core hardware) is certifiable as two notify bodies issued three certificates for railway systems last two years. All the technical justifications are in the 120 pages of the (not public) safety demonstration. The \CSSP\ seems competitive up to now as several contracts based on it have been won. However our best experts were involved in its development and first applications. The next systems based on it and developed by "more regular practitioners" will constitute the real test for its acceptance. Similarly the genericity of the platform will be assessed - implemented safety features and design degrees of freedom were designed to adapt to any "plausible" safety system. The extensions in Fig. \ref{process-future} have not been formally validated. The tools were developed mainly as proofs of concept, to assess if they comply to the 3-U rule: "useful, usable, used". In case of acceptance, stronger scientific work, drafted in this paper, will be required to either validate the existing translation principles or define new ones.

\section*{Acknowledgements}
The work and results described in this article were partly funded by BPI-France (Banque Publique d'Investissement) and Métropole Aix-Marseille as part of the project LCHIP (Low Cost High Integrity Platform) selected for the call AAP-21. This research was also partially funded by INES 2.0, FACEPE grant APQ-0399-1.03/17, CAPES grant 88887.136410/2017-00, and CNPq grant 465614/2014-0.
\bibliographystyle{splncs03}
\bibliography{biblio}
\end{document}